\documentclass[useAMS,usenatbib]{mn2e}
\usepackage{amssymb,amsmath,graphicx}
\usepackage{longtable}[1]
\usepackage{graphics}
\usepackage{threeparttable}
\usepackage{lscape}
\usepackage{color}
\usepackage{xspace}
\usepackage{url}

\newcommand{\Ha}{$\mathrm{H}\alpha$\xspace}
\newcommand{\NII}{$[\mathrm{N}\textsc{ii}]$\xspace}
\newcommand{\OII}{$[\mathrm{O}\textsc{ii}]$\xspace}
\newcommand{\NIIa}{$[\mathrm{N}\textsc{ii}]\,\lambda 6548$\xspace}
\newcommand{\NIIb}{$[\mathrm{N}\textsc{ii}]\,\lambda 6583$\xspace}
\newcommand{\WHa}{${W}_{\mathrm{H}\alpha}$\xspace}
\newcommand{\WNII}{${W}_{[\mathrm{N}\textsc{ii}]}$\xspace}
\newcommand{\ratio}{$[\mathrm{N}\textsc{ii}]\,/\,\mathrm{H}\alpha$\xspace}
\newcommand{\NIIratio}{$[\mathrm{N}\textsc{ii}]\,\lambda 6583 /
                        [\mathrm{N}\textsc{ii}]\,\lambda 6548$\xspace}

\title[OMEGA]
{OMEGA -- OSIRIS Mapping of Emission-line Galaxies in A901/2: I.-- Survey description, data analysis, and star formation and AGN activity in the highest density regions}
\author[Ana~L.~Chies-Santos~et~al.]
{Ana~L.~Chies-Santos$^{1,2}$\thanks{The first two authors have contributed equally to this paper.}\thanks{E-mail: ana.chies.santos@iag.usp.br}, Bruno~Rodr\'iguez~del~Pino$^{1,3}$\normalfont\textsuperscript{$\star$}, Alfonso~Arag\'on-Salamanca$^{1}$,
\newauthor
Steven~P.~Bamford$^{1}$, Meghan~E.~Gray$^{1}$, Christian Wolf$^{4}$, Asmus B\"ohm$^{5}$,
\newauthor
David T. Maltby$^{1}$, Irene Pintos-Castro$^{3,6,7,8}$, Miguel Sanch\'ez-Portal$^{8,9}$,
Tim Weinzirl$^{1}$\\
$^{1}$School of Physics and Astronomy, The University of Nottingham, University Park, Nottingham, NG7 2RD, UK\\
$^{2}$Departamento de Astronomia, Instituto de Astronomia, Geof\'isica e Ci\^encias Atmosf\'ericas, Universidade de S\~ao Paulo, S\~ao Paulo, SP, Brazil\\
$^{3}$Centro de Astrobiolog\'ia,INTA-CSIC, Madrid, Spain\\
$^{4}$Research School of Astronomy and Astrophysics, Australian National University, Cotter Road, Weston Creek, ACT 2611, Australia\\
$^{5}$Institute for Astro- and Particle Physics, University of Innsbruck, Technikerstr. 25/8, 6020 Innsbruck, Austria\\
$^{6}$Instituto de Astrof\'isica de Canarias, La Laguna, Tenerife, Spain\\
$^{7}$Universidad de La Laguna, Tenerife, Spain\\
$^{8}$ISDEFE, Madrid, Spain\\
$^{9}$European Space Astronomy Centre (ESAC)/ESA, P.O. Box 78, 28690 Villanueva de la Canada, Madrid, Spain\\
}

\begin{document}

\date{Accepted, 2 April 2015. Received, 9 February 2015; in original form, 21 November 2014}

\pagerange{\pageref{firstpage}--\pageref{lastpage}} \pubyear{2014}

\maketitle

\label{firstpage}

\begin{abstract}
We present an overview of and first results from the OMEGA survey: the OSIRIS Mapping of Emission-line Galaxies in the multi-cluster system A901/2.
The ultimate goal of this project is to study star formation and AGN activity across a broad range of environments at a single redshift.
Using the tuneable-filter mode of the OSIRIS instrument on GTC, we target \Ha and \NII emission lines over a $\sim \! 0.5 \times 0.5\;\mathrm{deg}^2$ region containing the $z \sim 0.167$ multi-cluster system A901/2.
In this paper we describe the design of the survey, the observations and the data analysis techniques developed. We then present early results from two OSIRIS pointings centred on the cores of the A901a and A902 clusters. AGN and star-forming (SF) objects are identified using the \ratio vs.\@ \WHa (WHAN) diagnostic diagram.
The AGN hosts are brighter, more massive, and possess earlier-type morphologies than SF galaxies.  Both populations tend to be located towards the outskirts of the high density regions we study.  The typical \Ha luminosity of these sources is significantly lower than that of field galaxies at similar redshifts, but greater than that found for A1689, a rich cluster at $z \sim 0.2$.
The \Ha luminosities of our objects translate into star-formation rates (SFRs) between $\sim \! 0.02$ and $6 \; M_{\sun}\,\mathrm{yr}^{-1}$.
Comparing the relationship between stellar mass and \Ha-derived SFR with that found in the field indicates a suppression of star formation in the cores of the clusters.
These findings agree with previous investigations of this multi-cluster structure, based on other star formation indicators, and demonstrate the power of tuneable filters for this kind of study. 
\end{abstract}


\section[]{Introduction}
Galaxies are not randomly distributed in the Universe:  while most are part of groups and clusters, they are also found in isolation or in the field. 
The observed properties of galaxies are
strongly dependent on the environment in which they reside (\citealt{dressler80}). In the local universe, galaxies in low-density environments are more likely to have spiral morphologies,
 blue colours and be forming stars, whereas those living in high-density regions are largely bulge-dominated, red 
and passive (\citealt*{balogh04}, \citealt*{bamford09}). These two populations may be connected by one or more evolutionary 
processes by which galaxies falling into denser environments have their star formation quenched and their
internal velocity dispersions increased. 

Both internal and external processes can drive this transformation. External processes such as ram-pressure 
stripping (\citealt*{gg72}, \citealt*{amb99}, \citealt*{bekki02}), tidal interactions (\citealt*{larson80}, \citealt*{bekki99})
galaxy harassment (\citealt*{moore96}), or mergers (\citealt*{barnes92}, \citealt*{bekki99}) can produce such morphological changes and cause the removal and/or 
consumption of gas, leading to the halt of star formation. 
On the other hand, processes related to internal galaxy properties, such as mass (\citealt{haines06}) and mass-dependent feedback mechanisms (\citealt*{bundy06}) can also be responsible for the quenching of star formation, 
as can feedback from supernovae and active galactic nuclei (AGN; \citealt{booth09}, \citealt{newton13}). 
In fact, environment also seems to 
be connected with the probability of a galaxy to host an AGN: while low-luminosity 
AGN do not show particular preference for any kind of environment (\citealt*{martini02}, \citealt*{miller03}, \citealt*{martini06}), luminous 
AGN are preferentially found in the field (\citealt*{k04}) and groups (\citealt*{popesso06}) rather than in clusters. \citet{peng10} and \cite{thomas10} showed 
that both stellar-mass and environmentally driven quenching of star formation are important in governing the evolution of a galaxy and that their effects can be separated 
from each other.

\cite{vogt04} and \cite{jaffe11} have shown that galaxies falling into clusters experience a removal of their gas, 
which in principle should lead to a gradual decline of the star formation activity. However, the average star formation rate (SFR) in star-forming (SF) galaxies 
remains roughly constant with environment, and it is only the fraction of star-forming galaxies that changes (\citealt*{balogh04}, 
\citealt*{verdugo08}, \citealt*{poggianti08}, \citealt*{bamford08}). The extent of ongoing star formation within a galaxy has also been shown 
to correlate with environment (\citealt*{kk04}, \citealt*{mow00}, \citealt*{bamford07}), implying a truncation of the star formation taking 
place first in the outer parts of the infalling galaxies. In fact, the last episode of star formation in cluster disc galaxies is seen to be concentrated towards the central regions 
(\citealt*{rodriguez14}, \citealt*{johnston14}). 

In order to understand the full degree of transformation and the relative role played by external and internal processes,
we selected the field of the A901/2 multi-cluster system at $z \sim 0.167$. This large structure has been the subject of study 
of the STAGES project (\citealt*{gray09}), where a wealth of data exists: an 80-orbit F606W \textit{HST}/ACS mosaic 
covering the full $\sim \! 0.5 \times 0.5\;\mathrm{deg}^2$ ($\sim \! 5 \times 5\;\mathrm{Mpc}^{2}$) span of the system, complemented by 
extensive multi-wavelength observations with \textit{XMM-Newton}, \textit{GALEX}, \textit{Spitzer}, 2dF, GMRT, Magellan, and the 17-band 
COMBO-17 photometric redshift survey (\citealt{wolf03}, \citealt{wolf04}). The broad range of galaxy environments and luminosities sampled by these observations
provide the perfect laboratory to study both obscured and unobscured star formation, stellar masses, AGN activity, and galaxy morphologies. 

Previous work in the A901/2 field has included studies of both star formation (\citealt{gray04}) and AGN activity and how they relate to galaxy environment.
\cite{wolf05} discovered red spirals in the system which suggested reduced current star formation rates in a large fraction of the galaxies.
\citet{gallazzi09} found an overall suppression in the fraction of star-forming
galaxies with density by combining the UV/optical SEDs from COMBO-17 and \textit{Spitzer} $24\,\micron$ photometry. The surviving star formation 
was shown to be obscured and hosted by a relatively large fraction of red spirals located primarily in the cluster infall regions (\citealt{wolf09}). In a more recent work, \citet{bosch13} find that this suppression of star formation is partially driven 
by ram-pressure stripping, which could lead to the production of red spirals and, ultimately, S0s.  Although the integrated star-formation properties
have yielded important results, the actual spatial scale of the star-forming regions and
how it is affected by the environment remains unknown. Likewise, initial attempts at quantifying the AGN activity in the field of A901/2 were made by \citet{gilmour07} and \citet{gallazzi09} using \textit{XMM-Newton} X-ray and  
\textit{Spitzer} $24\,\micron$ data, respectively. However, the lack of emission-line diagnostics means that the AGN census remains incomplete. 

In order to obtain spatially-resolved emission-line diagnostics, we have used the Gran Telescopio Canarias (GTC) Optical System for Imaging and low-Intermediate-Resolution Integrated Spectroscopy (OSIRIS) in tuneable filter mode to obtain very deep spatially-resolved emission-line images of the complete A901/2 STAGES field. We have targeted the \Ha ($\lambda_{0}=6563$\;\AA) and \NII ($\lambda_{0}=6583$\;\AA) lines. The former
is the best optical tracer of SFR (\citealt*{ken98}), whereas the combination of the two lines provides one of the most
reliable optical AGN diagnostics available (\citealt*{cid10}).

In this paper we present the design of the OSIRIS Mapping of Emission-line Galaxies in A901/2 (OMEGA) survey, the data reduction procedures adopted, and a first analysis of two of the twenty pointings, demonstrating some of the science that the survey can achieve.
OMEGA has been designed to address star formation and AGN activity as a function of galaxy properties and environment in A901/2. 
In the present work, we show a tentative census of AGN and star-forming galaxies in the highest-density cluster regions.
A full environmental analysis of both integrated and spatially-resolved galaxy properties will be presented in subsequent papers.

We adopt an $H_0=70\;\mathrm{km}\,\mathrm{s}^{-1}\,\mathrm{Mpc}^{-1}$, $\Omega_\mathrm{m}=0.3$ and $\Omega_{\Lambda}=0.7$ cosmology throughout.

\section[]{Description of the programme and design of the Observations}\label{description}

The OMEGA survey was designed to obtain very deep spatially-resolved emission-line images for the A901/2 multi-cluster system and, in turn, recover low-resolution spectra in a wavelength range that covers \Ha and \NII for the galaxies of the system.
Moreover, OMEGA is intended to sample the $16.0 \leq m_R \leq 23$ magnitude range. This roughly corresponds to stellar masses of $9.0 \leq \log{(M_{\star}/M_{\sun})} \leq 11.5$ at the cluster redshift, reaching lower SFRs than possible at high-$z$.  The survey is based on a 90\,h ESO/GTC Large Programme allocation (PI: A. Arag\'on-Salamanca).

In order to meet the objectives of OMEGA, we make use of the tuneable filter capabilities of the OSIRIS instrument (\citealt{cepa13a}, \citealt{cepa13b}), located at the Nasmyth-B focus of the 10.4\;m GTC, at the Roque de los Muchachos Observatory on La Palma.

Tuneable filter imaging is a method that lies between classical narrow-band imaging and spectroscopy, in terms of observing time costs and the quality of the information obtained.
Although the spectral resolution is relatively low, the wavelength sampling is quite sufficient for reliable star formation and AGN activity estimates for most of the cluster members.
Mapping the spectral region around \Ha, \NIIa and \NIIb for all the galaxies in the A901/2 system would be much more expensive with traditional spectroscopy.
A further advantage of tuneable filter imaging, with respect to slit spectroscopy, is the spatial information it provides. While long-slit spectroscopy normally only provides information along one of the axes of the galaxies (usually chosen to be the major axis), tuneable filter imaging provides 2D maps.
The tuneable filter technique has not been much employed previously in this field. A heroic attempt, includes the CADIS survey of emission line galaxies measuring SFRs between $z=0.25$ and $z=1.2$ (\citealt{hippelein03}).

Our increased spectral resolution over standard narrow-band \Ha imaging surveys (see, e.g., \citealt{ken92}, \citealt{gal97}) has the advantages of avoiding contamination from the \NII line in the star formation estimates and enabling AGN to be identified (\citealt{cid10}).
Finally, while integral field spectrograph (IFS) surveys, such as CALIFA and MaNGA, also provide spatial information, their fields of view are much smaller than that of OSIRIS. Even large IFSs -- such as MUSE, recently installed on the VLT -- have a field of view that is $\sim$50 times smaller than that of OSIRIS.  The full unvignetted field-of-view (FOV) of OSIRIS is $7.8 \times 7.8\;\mathrm{arcmin}^2$, imaged using two $2048 \times 4096$ Marconi CCDs with a $9.4\;\mathrm{arcsec}$ gap between them. The 
plate scale, using $2 \times 2$ pixel binning, is $0.25\;\mathrm{arcsec}\,\mathrm{pixel}^{-1}$.  
We can therefore obtain spectral and spatial information for a much larger number of galaxies, and thus gain a much more complete picture, than is possible with current IFSs.

The OSIRIS tuneable filter system consists of two reflecting plates (a Fabry-P\'erot etalon) working in a collimated beam.  The resulting interference allows only specific narrow wavelength ranges to be transmitted.
One can tune the properties of the filter by modifying the separation between the optical plates of the device. Standard narrow-band order-sorting filters (OSF) are used to isolate specific interference orders. Reliable observations are limited to a circular field of view of radius $4\;\mathrm{arcmin}$, within which they are assured to be free from contamination by other orders. OSIRIS has two different tuneable filters, one for the blue and one for the red wavelength range. We use the latter, which is capable of tuning filters to wavelengths in the 6510--9350\;\AA\ range with a spectral resolution of 14\AA\ FWHM. 

A given radius from the optical centre in the image plane corresponds to a specific angle in the incoming collimated beam, which is incident on the Fabry-Perot etalon. As the transmission wavelength of the interferometer depends upon incidence angle, the wavelength of the transmitted light varies as a function of position in the image.
The result is a circular pattern with rings of constant effective wavelength, $\lambda$, which depends on the radius, $r$, from the optical centre. In order to achieve the best accuracy in the estimation of the wavelength we use the following expression (\citealt{gonzalez14}):
\begin{equation}
\lambda=\lambda_{0} - 5.04\,r^{2}\ + a_3(\lambda)r^{3},
\label{lambdavar}
\end{equation}
where
\begin{equation}
a_3(\lambda)=6.0396 - 1.5698 \times 10^{-3}\lambda + 1.0024 \times 10^{-7}\lambda^{2},
\label{lambdavar2}
\end{equation}
where $\lambda$ is in \AA\ and $\lambda_{0}$ is the effective wavelength at the optical centre.
Since the useable FOV is limited to a radius of $4\;\mathrm{arcmin}$, the maximum variation in the effective wavelength is 80\;\AA.

\begin{figure*}
\begin{center}
\includegraphics[width=15cm]{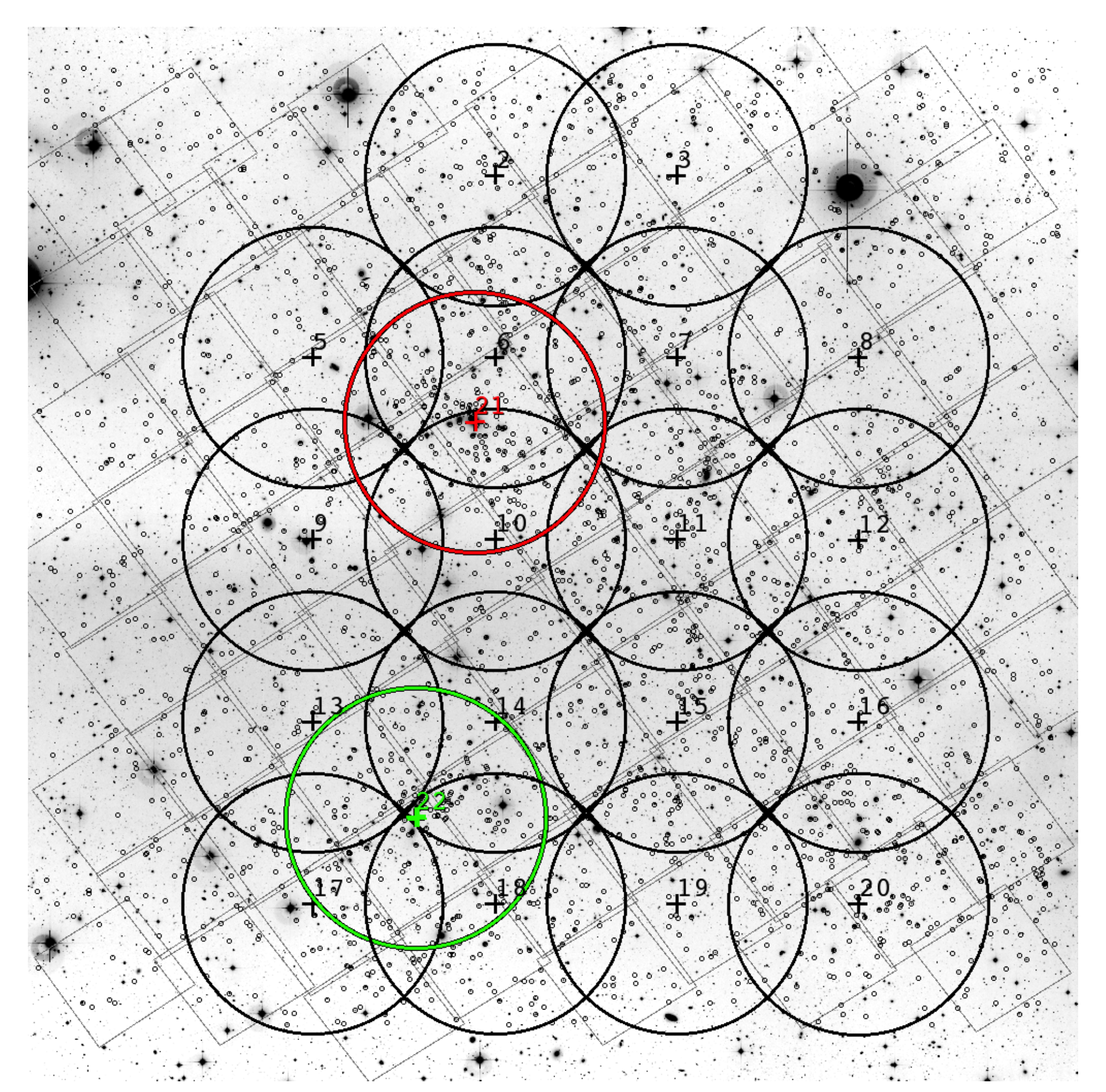}
\caption{The OSIRIS observed fields (labeled 2--3, 5--22), represented as circles of radius $4\;\mathrm{arcmin}$, overlaid on top of the footprint of the 80-tile STAGES HST/ACS mosaic and a COMBO-17 $R$-band image. The very small black circles mark the positions of cluster galaxies with $m_R<23$, where cluster membership is defined as in \citet{gray09} using the COMBO-17 photometric redshifts.
}
\label{fieldlayout}
\end{center}
\end{figure*}

In Fig.~\ref{fieldlayout} we show the OSIRIS observed fields as circles $8\;\mathrm{arcmin}$ in diameter, overlaid on top of the footprint of the 80-tile STAGES \textit{HST}/ACS mosaic (\citealt{gray09}) and ground-based COMBO-17 $R$-band image (\citealt{wolf04}). Cluster galaxies with $m_R<23.5$ are also plotted. Given the available observing time, the initial plan was to tile the field-of-view uniformly in a $4 \times 5$ grid. However, the North-East and North-West tiles contain bright stars that would have produced reflection ghosts, seriously compromising the quality of the data for these tiles. For this reason, we decided to re-locate these two pointings to cover the regions with the highest galaxy densities, which correspond to Fields 21 (A901a) and 22 (A902). These fields are referred to as F21 and F22 respectively. 

\begin{figure}
\begin{center}
\includegraphics[width=0.48\textwidth]{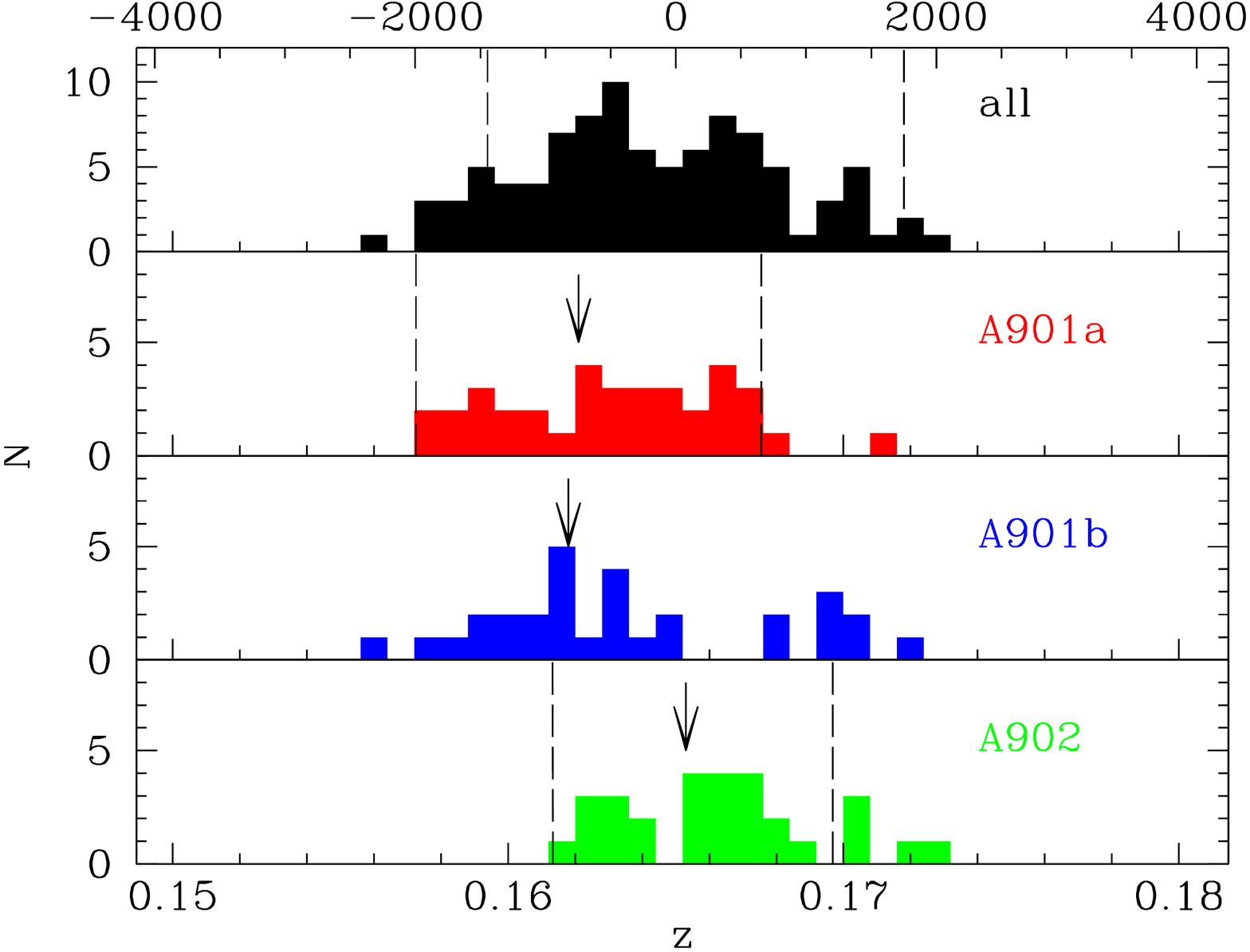}
\caption{Spectroscopically determined redshift distribution of galaxies
within $1.2\;\mathrm{Mpc}$ diameter apertures around the three main cluster cores, from a limited spectroscopic survey of the $\sim \! 300$ brightest galaxies in the region, taken with the AAT 2dF instrument. Arrows mark the positions of the brightest cluster galaxies. Dashed lines indicate the velocity range probed by OMEGA:  the top histogram indicates the limits of the survey proper (fields 2, 3 and 5--20), while the ranges for the two pointed observations of the A901a and A902 cluster cores  presented in this paper (fields 21 and 22 respectively) are specially tuned (cf. Table~\ref{OBS}).}
\label{redshifts}
\end{center}
\end{figure}

Fig.~\ref{redshifts} shows the spectroscopically determined redshift distribution of galaxies within $1.2$\;Mpc diameter apertures centred upon the three main cluster cores.  The redshifts were drawn from a spectroscopic survey of the 300 brightest cluster galaxies in the region obtained with the AAT 2dF spectrograph (Gray et al., in prep.). The dashed lines indicate the redshift range probed by our observations.
Within these redshift ranges we target both the \Ha and \NIIb emission lines.  The $z$ ranges are slightly narrower for the two pointed observations of the A901a and A902 regions (F21 and F22).

\begin{figure*}
\begin{center}
\includegraphics[width=\textwidth]{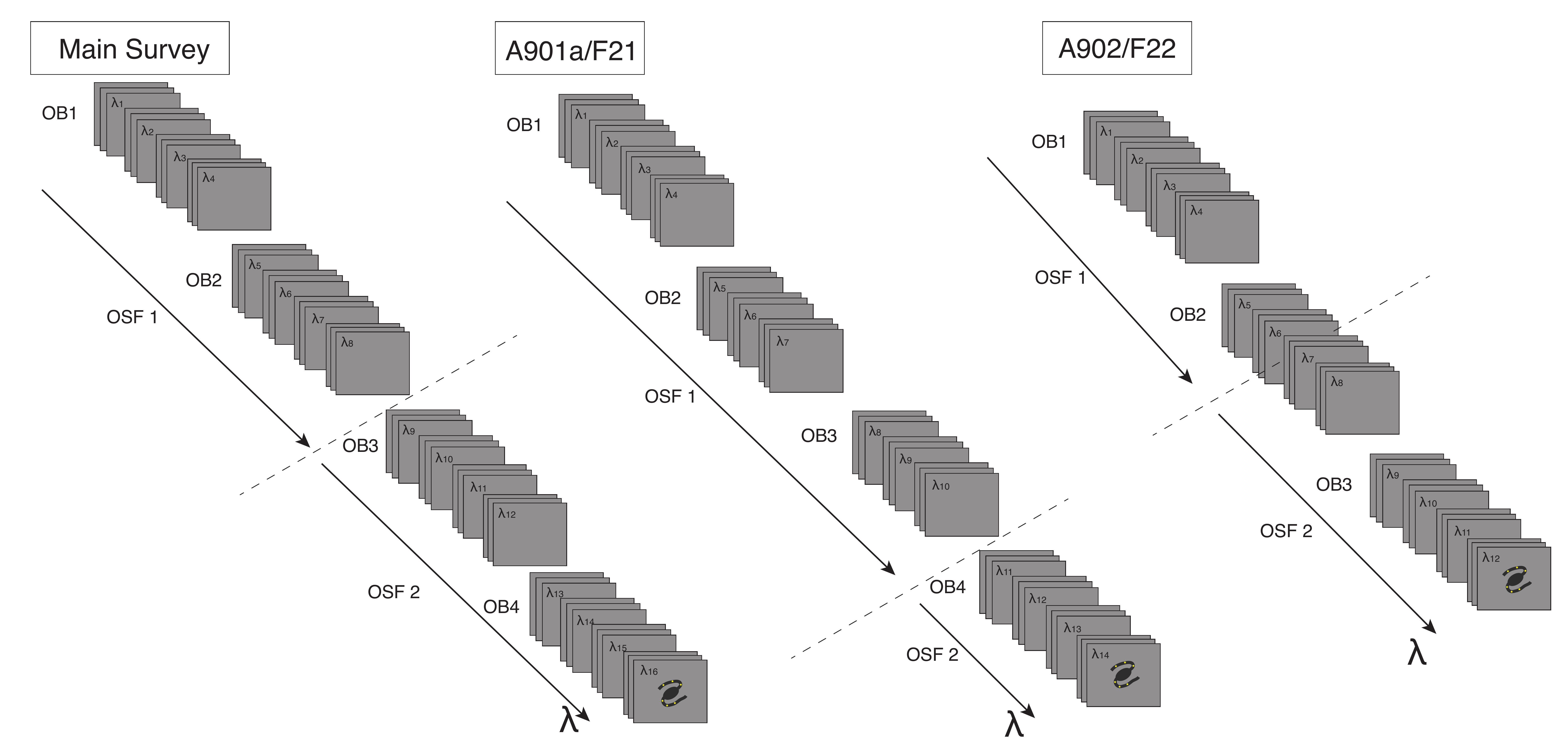}
\caption{A cartoon illustrating the design of the observations for the main survey and the two cluster core fields (F21 and F22). Note that this illustration is valid in the case of a single galaxy, for which the wavelength is almost constant (for a given image there is wavelength variation across the field-of-view, see Eq.~\ref{lambdavar} and Fig.~\ref{lambda_dependence}). A given galaxy is observed using a series of wavelength setups $\lambda_i$ and a 3-position dithered pattern. In each dithered position the wavelength changes slightly because the distance to the centre of the field changes. The order-sorting filter (OSF) separates the different tuned wavelengths. The observations are executed in `Observing Blocks' (OBs).    
See http://www.nottingham.ac.uk/astronomy/omega/videos/ for videos with successive frames of a given galaxy.
}
\label{datacubes}
\end{center}
\end{figure*}

\begin{figure}
\begin{center}
\includegraphics[width=0.48\textwidth,clip,trim = 80 25 50 25]{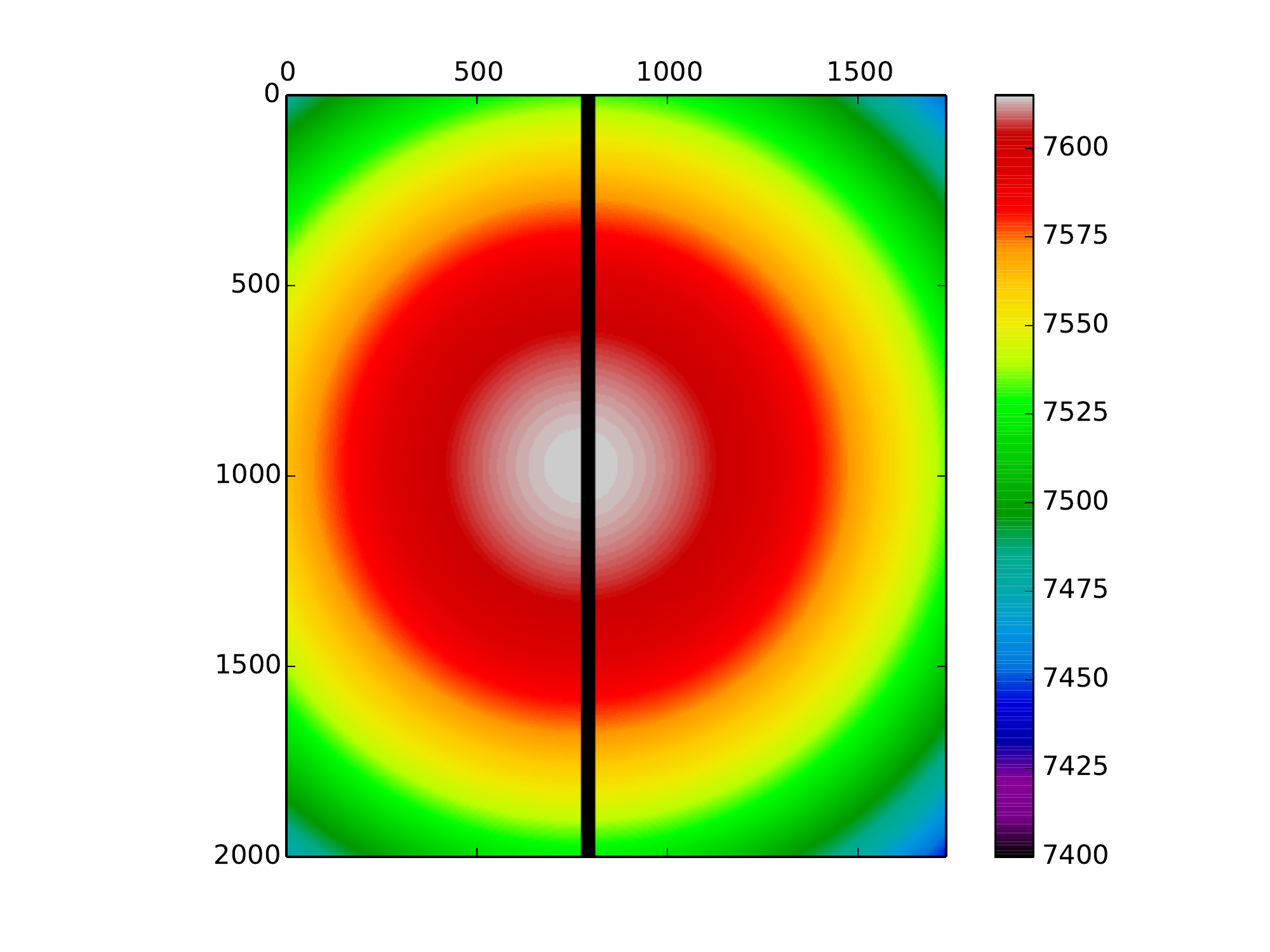}
\caption{A mock reduced image with $\lambda_\mathrm{c}=7615$\;\AA, illustrating the radial wavelength dependence according to the colour bar (labelled by wavelength in \AA).  The vertical line in the middle represents the $9.4\;\mathrm{arcsec}$ gap between CCD1 and CCD2 and the axis are in pixels.}
\label{lambda_dependence}
\end{center}
\end{figure}

For an optimal deblending of \Ha and \NII, the tuneable filter Full Width at Half Maximum (FWHM) bandwidth is set to 14\;\AA, and the spacing between successive wavelengths to 7\;\AA. Therefore, in order to cover the desired wavelength range in F21 and F22 we require 14 and 12 wavelength settings, respectively, whereas for the rest of the fields we need 16 wavelength settings (see Figs.~\ref{redshifts}~and~\ref{datacubes}, and Table~\ref{OBS}).  Moreover, for each wavelength setting, 3 dithered images are taken typically with a $\sim40$\,pix (10$\arcsec$) separation. This increases our number of images (and individual wavelengths sampled) by a factor of three and ensures that the inter-CCD gap is imaged. A total of 42 and 36 images have been obtained for Fields 21 and 22 respectively, and 48 for the remaining 18 fields. Note however, that the wavelength varies across the field, according to Eq.~\ref{lambdavar}. In Fig.~\ref{lambda_dependence} we show an illustration of this dependence. 

Since the wavelength varies across the FOV, the central wavelength $\lambda_\mathrm{c}$ of each wavelength setting needs to be optimised in such a way that the desired wavelength range is covered for as many galaxies as possible.
The average wavelength $\lambda_\mathrm{ave}$ of each setting was calculated as the mean $\lambda$ within the central $5.6\;\mathrm{arcmin}$ diameter circle
of each pointing (the approximate area uniquely covered by one tile).
For fields F2 to F20, $7609\,\le \lambda_\mathrm{ave} (\textrm{\AA})\le 7714$,
ensuring that both \Ha and \NII are observed in the $0.1594 \le z \le 0.1718$ redshift ranges. This corresponds to rest-frame velocities in the range $-1447 \le v(\textrm{km/s}) \le 1751$, where $v=0$ corresponds to $z=0.167$, the average redshift of the structure.
For field F21, $7595\, \le \lambda_\mathrm{ave}(\textrm{\AA}) \le 7686$, and
for F22, $7623\, \le \lambda_\mathrm{ave}(\textrm{\AA}) \le 7700$,
thus covering a slightly narrower redshift/velocity range for these two fields (see Fig.~\ref{redshifts} and Table~\ref{OBS} for details).
Note that these are average values: the exact wavelength range covered for each galaxy depends on its position with respect to the centre of the pointing. Moreover, given the significant tile overlap, a sizeable fraction of the galaxies will be observed in more than one pointing, increasing the available wavelength range and the number of independent wavelength samples obtained. 

A summary of the observational setup is given in Table~\ref{OBS}.
The observations were executed in `Observing Blocks' (OBs), as shown in Table~\ref{OBS} and illustrated in Fig.~\ref{datacubes}. They were carried out during the first halves of 2012 and 2013, and completed in February 2014.

\begin{table*}
\caption{Summary of the Observations.}
\label{OBS}
\begin{tabular}{lcccccccc}
\hline
OBs (Main Survey) & OBs (A901a/F21) & OBs (A902/F22) & $\lambda_\mathrm{c}$ (\AA) & $\lambda_\mathrm{ave}$ (\AA) & Order Sorting Filter & Exp.\@ Time (s)\\
\hline
 &1 &  &7615 & 7595 & f754/50&  3$\times$200\\ 
 &1 &  &7622 & 7602 & f754/50&  3$\times$200\\    
1&1 &  &7629 & 7609 & f754/50&  3$\times$200\\
1&1 &  &7636 & 7616 & f754/50&  3$\times$200\\
1&2 &1 &7643 & 7623 & f754/50&  3$\times$200\\
1&2 &1 &7650 & 7630 & f754/50&  3$\times$200\\
2&2 &1 &7657 & 7637 & f754/50&  3$\times$200\\
2&3 &1 &7664 & 7644 & f754/50&  3$\times$200\\
2&3 &2 &7671 & 7651 & f754/50&  3$\times$200\\
2&3 &2 &7678 & 7658 & f754/50&  3$\times$200\\
3&4 &2 &7685 & 7665 & f770/50&  3$\times$200\\
3&4 &2 &7692 & 7672 & f770/50&  3$\times$200\\
3&4 &3 &7699 & 7679 & f770/50&  3$\times$200\\
3&4 &3 &7706 & 7686 & f770/50&  3$\times$200\\
4&  &3 &7713 & 7693 & f770/50&  3$\times$200\\
4&  &3 &7720 & 7700 & f770/50&  3$\times$200\\
4&  &  &7727 & 7707 & f770/50&  3$\times$200\\
4&  &  &7734 & 7714 & f770/50&  3$\times$200\\

\hline
\end{tabular}
\end{table*}

\section[]{Data Reduction}
In this section we describe the data reduction methods we have developed for the OMEGA data.

\begin{figure}
\begin{center}
\includegraphics[width=8cm]{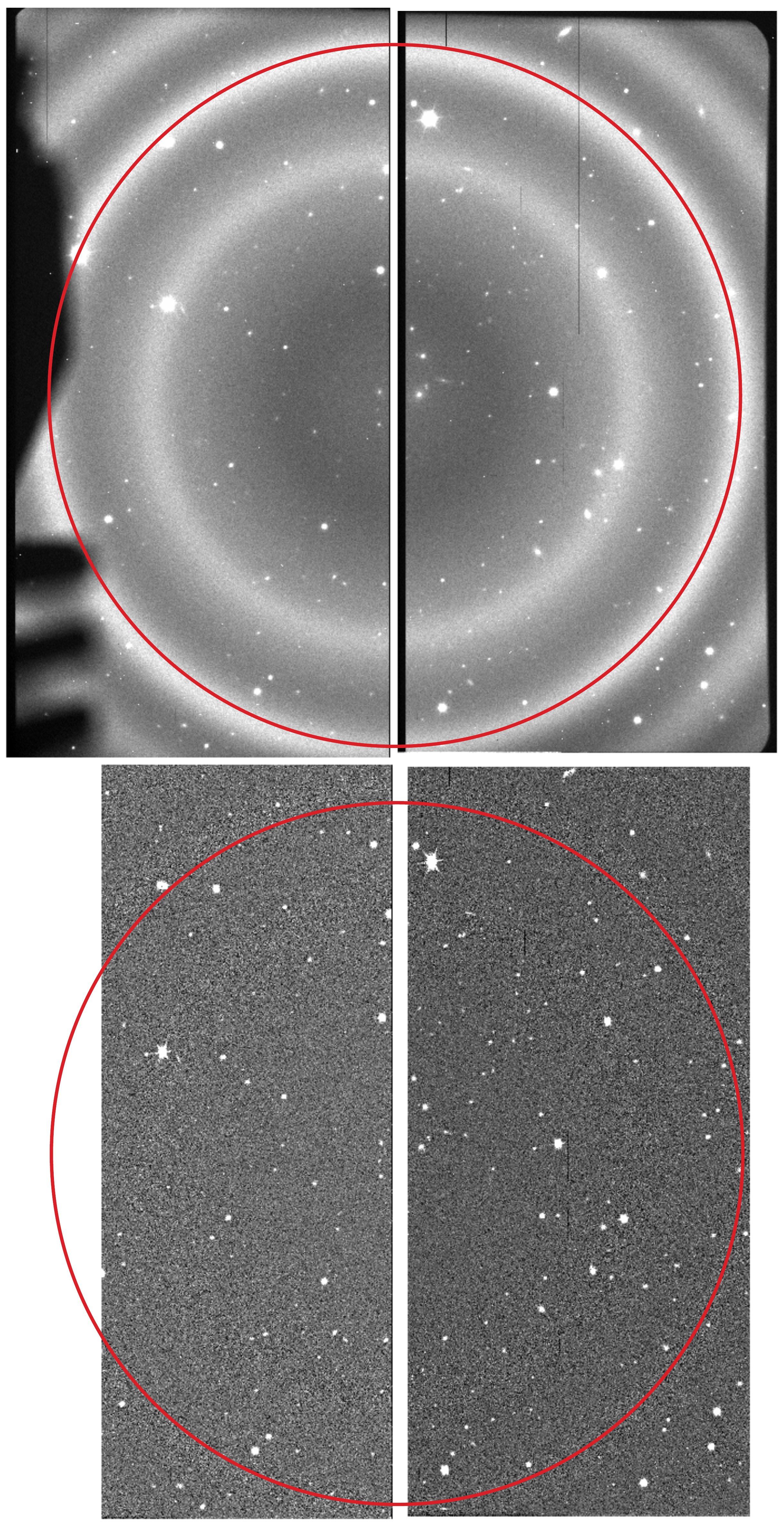}
\caption{Example of the first image of F21 with $\lambda_\mathrm{c}=7615$\;\AA\@. \emph{Top panel:} raw image. \emph{Bottom panel:} after OOPs processing. The red circle has a radius of $4\;\mathrm{arcmin}$, within which the observations are free from contamination by other interference orders.}
\label{oops}
\end{center}
\end{figure}

\subsection[]{Imaging Data Reduction}
The data reduction was carried out with the Osiris Offline Pipeline software \citep[OOPs; ][]{oops}. 
The OOPs reduction consists of two steps. In the first step we prepare the calibration files `masterbias' and `masterflat' combining the individual bias and flatfield frames taken at the telescope.  
Flatfields were obtained using dome flats with the filter tuned to the same wavelength as the science observations. Usually 3--5 flat frames were taken each night at each wavelength.
In the second step we reduce the science frames: we remove the overscan and subtract the bias from the science frames and divide these frames by the masterflat. The vignetted regions of the images are also trimmed at this stage. 

The final step carried out within OOPs is the sky subtraction.
Due to the wavelength variation across the FOV, sky OH emission lines appear as rings (see top panel of Fig.~\ref{oops}). To correct for these features, each science exposure is artificially dithered on a 3 by 3 grid. The dither step size (a few arcseconds) is larger than the vast majority of the sources, but small compared to the ring features.  
Therefore, median-combining the dithered images leaves only the sky rings.  These can then be subtracted from the original image. 
Fig.~\ref{oops} presents the first image of Field 21, with $\lambda_\mathrm{c}=7615$\;\AA. The upper panel shows the raw image while the bottom panel displays the final output from OOPs.

As mentioned in Section~\ref{description}, we obtain three dithered exposures for each $\lambda_\mathrm{c}$ (see Fig.~\ref{datacubes}). This dithering introduces small shifts in the wavelength corresponding to each pixel.  As we wish to retain this information, to improve our spectral sampling, we do not coadd the dithered images.  To correct for cosmic rays, we therefore employ a similar routine to \textsc{qzap} in \textsc{iraf} applying a $3\sigma$ clipping with a fluxratio  of $0.1$ (comparison value between the flux in the surrounding pixels and the flux in  the candidate one to check whether the pixel has been hit by a cosmic ray).

Bad pixels are corrected using the \textsc{iraf} task \textsc{fixpix}. Astrometry solutions are found using \textsc{ccmap} and \textsc{ccsetwcs} with the COMBO-17 $R$-band image as reference. In order to avoid unnecessary distortions, we work with the two separate CCDs of OSIRIS throughout the analysis, obtaining astrometric solutions with rms accuracies $\leq 0.1\;\mathrm{arcsec}$.

\subsection[]{Wavelength Calibration}

The nature of the observations mean that the wavelength of a given object will depend on the central wavelength of each image as well as its distance to the optical centre. Both parameters change from exposure to exposure. The variation of wavelength across the field-of-view for a given central wavelength is shown in Fig.~\ref{lambda_dependence}.

The wavelength calibration of the OSIRIS tuneable filter has been a question of debate (\citealt{ma11}) because its accuracy is crucial for studies of extended sources, as well as for estimating redshifts of both point-like and extended sources. Fortunately, the sizes of our objects are small (the largest objects in F21 and F22 have a radius of $\sim \! 7\;\mathrm{arcsec}$), and hence the wavelength changes are negligible across individual objects in the OMEGA sample (at most $\sim2$\AA, i.e., much smaller than the spectral resolution, 14\AA\,FWHM).
The general form of the $\lambda$ calibration follows from Eq.\,\ref{lambdavar}. The optical centre is located on pixel $X_0=1051$ and $Y_0=976$ (CCD1) and $X_0=-10$ and $Y_0= 976$ (CCD2) of the raw images, according to the OSIRIS website.\footnote{http://www.gtc.iac.es/instruments/osiris/} This corresponds to $X_0=772$, $Y_0=976$ (CCD1) and $X_0=-35$ and $Y_0=976$ (CCD2) on the images output from OOPs.  
Although there are other optical centres listed in the literature (\citealt{ma11} and the OSIRIS manual), the change they introduce is smaller than $0.2$ per cent in wavelength.
Using either of these values makes no difference for the purposes of the current paper. 
Here we are interested primarily in the measurement of the \Ha and \NII line fluxes, and not in the determination of accurate redshifts. However, in future papers, the combination of the images for all 20 OMEGA fields will allow us to evaluate the accuracy of the wavelength calibration using a very large number of galaxies in the overlap regions (see Fig.~\ref{fieldlayout}) that are observed two or even three times. Furthermore, the available AAT 2dF spectroscopic redshifts will provide an external cross-check.

\subsection[]{Flux calibration and zero-point estimation}
Different image sets (OBs) for a single field are not necessarily observed on the same night or with the same observing conditions. In order to perform consistent photometry, we therefore need to correct for differences in the seeing conditions, transparency and airmass. 

To match the seeing conditions our procedure is as follows: we measure the PSF in all the images using unsaturated stars with $m_{R,\mathrm{COMBO-17}}<19$ (from the STAGES catalogue of \citealt{gray09}).  We then convolve each image with a Gaussian function of the required width to produce images with the same PSF.  From now on we use images matched to the worst seeing, which is 1.2\,arcsec. These operations were performed using the \textsc{iraf} tasks \textsc{psfmeasure} and \textsc{gaussian}. 

We perform photometric measurements using \textsc{phot} within \textsc{iraf} on the stars in each individual OSIRIS image (typically, 15--25 stars per field) using an aperture of $2\;\mathrm{arcsec}$ ($8\;\mathrm{pixels}$) radius, sufficient to measure the total flux for these point sources. Because the wavelength at which each star is observed varies for different images, we obtain a wavelength-dependent zero-point $\mathit{zp}(\lambda)$ function,
\begin{equation}
\label{zp}
\mathit{zp}(\lambda)=m_{\star\,\mathrm{W753f}} - m_{\star\,\mathrm{OSIRIS}}\;,
\end{equation}
where $m_{\star\,\mathrm{W753f}}$ is the standard magnitude of that star, defined below, and $m_{\star\,\mathrm{OSIRIS}}$ is the instrumental magnitude of the star measured in \textsc{phot}.

\begin{figure}
\begin{center}
\includegraphics[width=.48\textwidth]{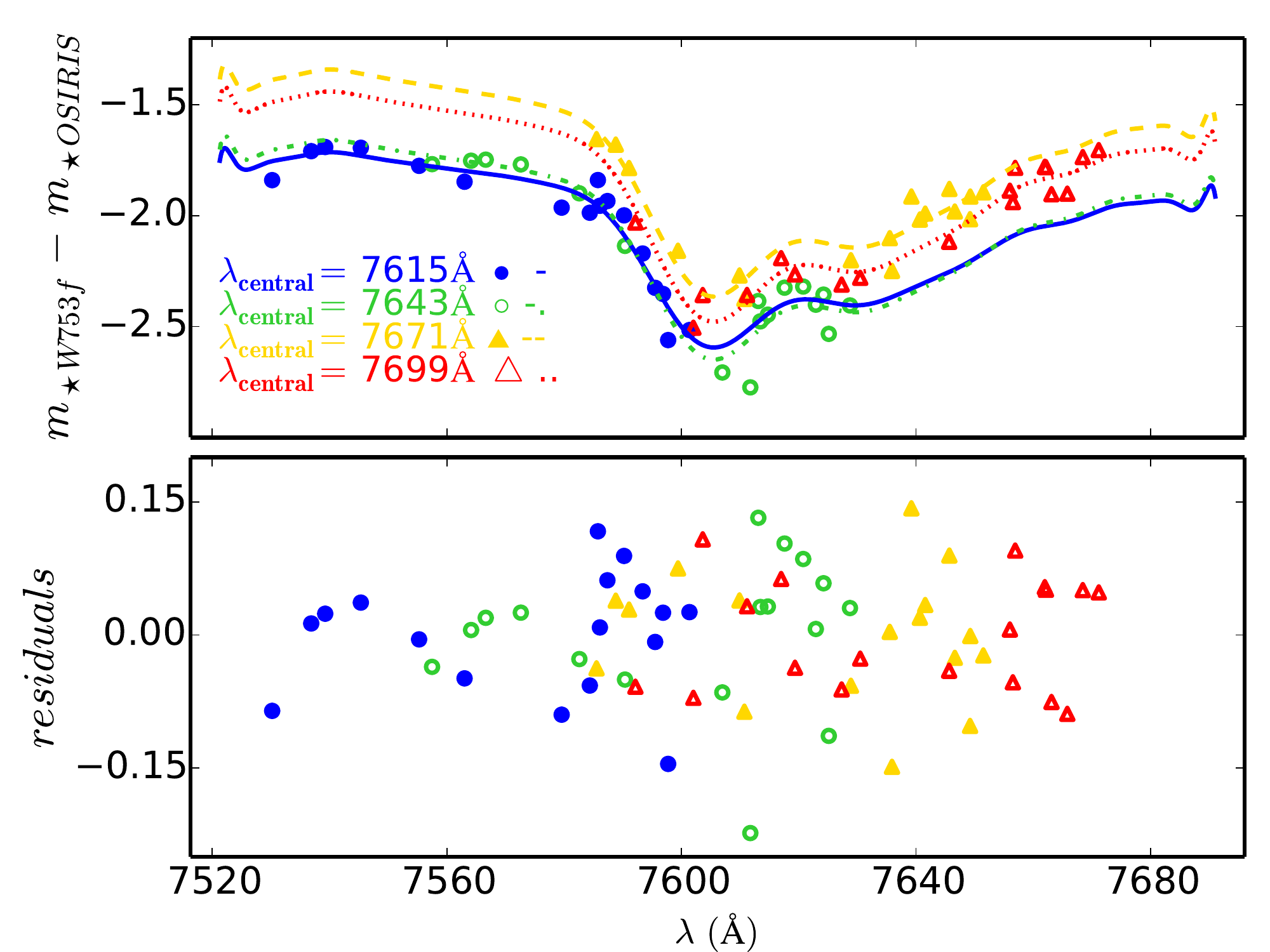}
\caption{\emph{Top panel:} The wavelength-dependent zero-point, $\mathit{zp}(\lambda)=m_{\star\,\mathrm{OSIRIS}}-m_{\star\,\mathrm{W753f}}$, for 4 different dithered images in Field 21. Each set of $\sim \! 25$ stars is shown with a different colour and symbol. We only show 4 of the total 42 images for clarity. In order to model the effect of the telluric line, we fit a polynomial. The best solution for each image set is shown by the lines. \emph{Bottom panel:} The residuals with respect to the polynomial fits.}
\label{zeropoint}
\end{center}
\end{figure}

To calibrate our fluxes we use the Vega magnitudes of the same stars in the closest COMBO-17 filter, $\mathrm{W753f}$, as standard magnitudes. This filter samples the range 7440--7620\;\AA.
Vega magnitudes were calculated from the photon fluxes tabulated in the COMBO-17 catalogue.
Our zero-point functions are shown in Fig.~\ref{zeropoint}.  Because there is a strong absorption feature caused by a double telluric sky line at $\lambda=7633$\;\AA, the zero-point function has a strong wavelength dependency. 

In order to correct for the effect of the atmospheric telluric absorption, we fit this feature with a series of Chebyshev polynomials. The shape of the absorption feature is constant, except for a scaling factor and offset that vary with the observing conditions.  We therefore fit all the zero-point functions simultaneously, to obtain a general set of polynomial coefficients, together with an offset and scaling factor for each set of dithered images (14 in the case of Field 21, and 12 for Field 22).  We use a similar fitting technique to that described later, in Sec.~\ref{sec:fitting}. In Fig.~\ref{zeropoint} we show the zero-points, as well as our fits and their residuals, as a function of wavelength. For clarity, we only show these for 4 images from the set of 14 dithered images. 

The scatter in the zero-point introduces additional wavelength-dependent errors that must be incorporated in our photometric uncertainties. We estimate these by taking a $3\sigma$--clipped standard deviation of the residuals in bins of 10\;\AA.
These zero-point errors are found to be always $\la 0.1$\;mag.

\subsection[] {Galaxy photometry}

We perform aperture photometry on the galaxies with \textsc{phot}.  We measure each galaxy using two apertures: a PSF-matched aperture of radius $1.2\;\mathrm{arcsec}$ ($5\;\mathrm{pixels}$, from now on referred to as $R_{\rm PSF}$) and a total aperture (essentially the Kron radius; \citealt{ba96}), with $R_{\rm tot} = 2.5 \times \mathit{SMA}$.
Here $\mathit{SMA}$ is the second central moment of the light distribution, derived from ACS images with \textsc{SExtractor} and contained within the STAGES catalogue (\citealt{gray09}).
We use these two different aperture radii as appropriate throughout the paper. $R_{\rm PSF}$ provides information about the nucleus of a given galaxy, while $R_{\rm tot}$ averages over its full extent.  Fig.~\ref{two_spectra} shows the spectra constructed for one of the galaxies in F21 using both $R_{\rm PSF}$ or $R_{\rm tot}$ apertures.

Since at different wavelengths the galaxies' light distributions may change, we do not re-centre the aperture for each measurement, but instead fix the aperture position to the RA and Dec from the STAGES catalogue of \cite{gray09}.
The flux calibration is then applied using the zero-point $\mathit{zp}(\lambda)$ previously obtained via Eq.~\ref{zp}:
\begin{equation}
\label{magcalib}
m_\mathrm{calib}(\lambda)=m_\mathrm{OSIRIS}(\lambda) - \mathit{zp}(\lambda) \;. 
\end{equation}
When working in fluxes this becomes
\begin{equation}
\label{fluxcalib}
F_\mathrm{calib}(\lambda)=F_\mathrm{OSIRIS}(\lambda) \times 10^{\mathit{zp}(\lambda)/2.5} \;, 
\end{equation}
where $F_\mathrm{calib}(\lambda)$ is our calibrated flux in ${\rm ergs}\,{\rm cm}^{-2}\,{\rm s}^{-1}\,\textrm{\AA}^{-1}$ and $F_\mathrm{OSIRIS}(\lambda)$ is the flux measured with \textsc{phot}.

\subsection[] {Building the spectra}

\begin{figure*}
\begin{center}
\includegraphics[width=15cm]{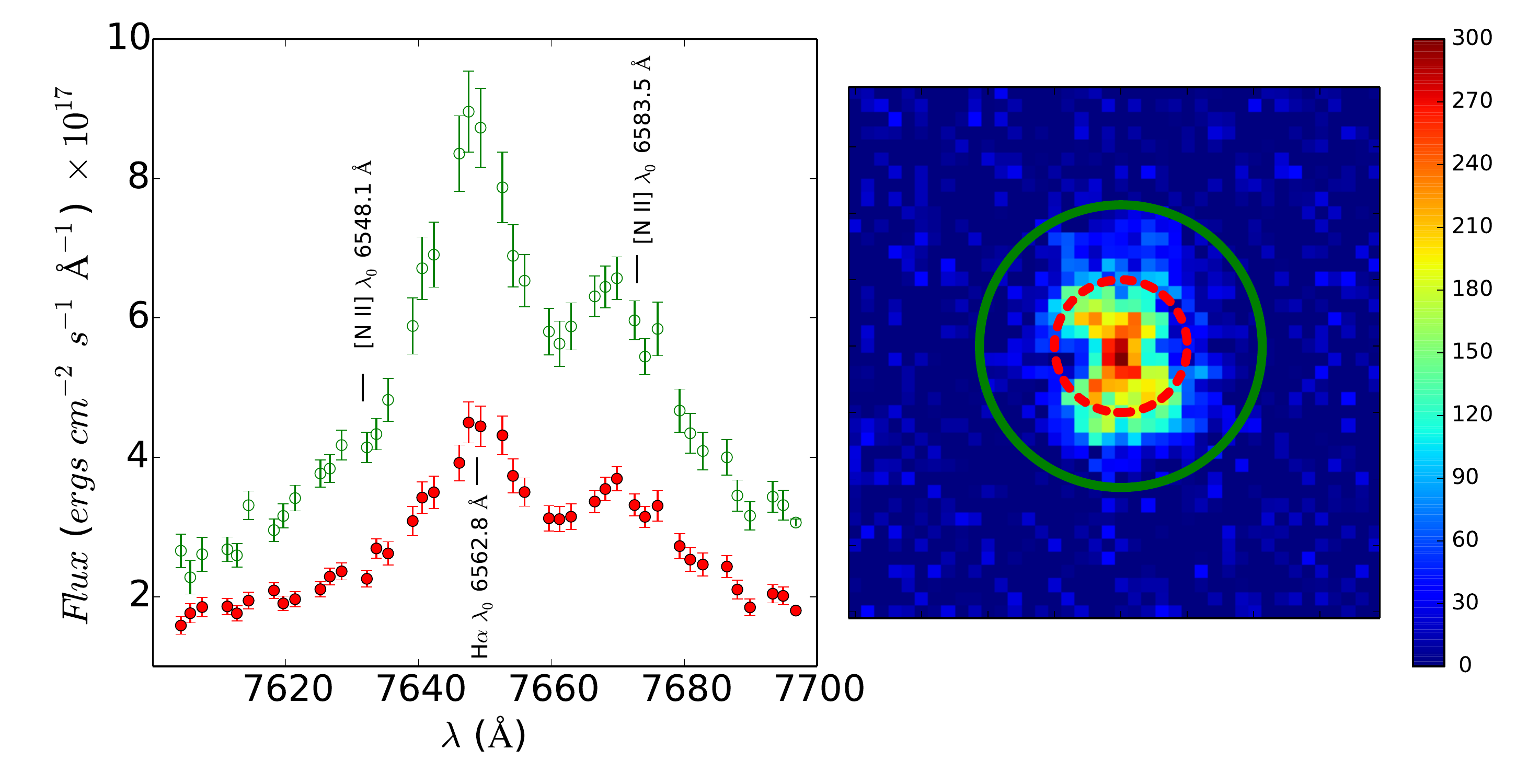}
\caption{\textit{Left panel:} The flux-calibrated spectra for the central (filled red circles) and total (open green circles) apertures of the bright ($m_R\sim18.7$) galaxy with STAGES ID 42713. The locations of \Ha ($\lambda_0 = 6562.8\,\textrm{\AA}$) and the two \NII lines ($\lambda_0 = 6548.1\,\textrm{\AA}$ and $6583.5\,\textrm{\AA}$) are indicated. \textit{Right panel:} Image of galaxy 42713 at $\lambda=7646\,\textrm{\AA}$, with the total and PSF apertures overlaid as solid green and dashed red lines, respectively. This image has not been convolved to the standard PSF
although the fluxes are actually measured on the convolved images. The image is 40 pixels on a side, which corresponds to $10\;\mathrm{arcsec}$ or $28\;\mathrm{kpc}$ at $z=0.167$. The colour bar indicates the flux in counts.}
\label{two_spectra}
\end{center}
\end{figure*}

After the flux and wavelength calibrations are performed, we are able to construct a spectrum for each of the galaxies in the observed fields. An example of such a spectrum can be found in the left panel of Fig.~\ref{two_spectra}, where we indicate the location of \Ha and the two \NII lines  with their respective rest-frame wavelengths.
By design, the wavelength range of the spectra for cluster members should cover both \Ha and 
\NIIb, which will appear as emission lines in galaxies hosting star formation and/or AGN activity. 
In addition to finding spectra of galaxies with clear emission lines, we also expect to obtain spectra of galaxies with weak or absent emission (or absorption in \Ha) lines.

\section[]{Analysis}

\subsection[]{Spectral fitting}
\label{sec:fitting}

Our goal is to measure fluxes for \Ha as well as the brightest \NII line at 6583\;\AA.
We also wish to obtain reliable constraints for the continuum, in order to measure the equivalent widths (\WHa and \WNII) of these lines. Due to the limited wavelength coverage and the moderate spectral resolution and sampling of our spectra, this exercise becomes very challenging for weak emission-line galaxies (ELGs).

To account for all the features present in our spectra, we fit them with a composite of three Gaussian functions and a linear continuum. Since the ratio of the fluxes of the two nitrogen lines (\NIIratio) is fixed by atomic physics to a value of $3.06$ (determined by the ratio between the transition probabilities of each line,  \citealt*{oster89}), we only need one parameter to fit both their fluxes. We therefore fit, at most, six parameters:
continuum intercept, continuum slope, redshift, flux of \Ha, width of \Ha, and flux of \NIIb.

In order to exploit all the information contained in our data, the fitting procedure must evaluate all  possible combinations of the parameters in a thorough yet efficient way. Conventional fitting techniques, such as Levenberg-Marquardt optimisation, can be sensitive to the initial values, and can become trapped in local maxima in the parameter likelihood space. These problems can be ameliorated  by using Markov Chain Monte Carlo (MCMC) techniques. In particular, we use the software \textsc{emcee} (\citealt{mcmcham}), which implements the Affine Invariant MCMC Ensemble sampler of \cite{gw10}.

MCMC methods aim to draw samples from the posterior probability density distribution, which describes how likely each possible set of model parameters is given the data.  From Bayes Theorem, the posterior is given by the product of the prior distribution (which can be thought of as pre-existing constraints on the parameter values) and the likelihood function (which describes how closely the data correspond to the model, for a given set of parameters). MCMC algorithms explore the parameter space along a chain.  At each step, new values for the parameters are proposed.  These are accepted or rejected, depending on the probability of the proposed values relative to that of the current parameter set. If the probability is higher, then the proposed parameters will be adopted. However, if the probability is lower, there remains some chance of the new values being accepted.  In this way, the parameter space is thoroughly explored, and ultimately sampled in proportion to the posterior probability density.

The Affine Invariant MCMC Ensemble sampler uses multiple `walkers' to produce a set of chains. These are started at different points of the parameter space. Each walker explores following a proposal distribution that depends on the current position of the rest of the walkers. In this way the parameter space can be mapped more efficiently than with the original Metropolis-Hastings algorithm. We also make use of parallel tempering, in which additional sets of walkers take significantly larger steps than the standard set.  This eases communication between different maxima in the case of multimodal posterior distributions and generally speeds up convergence. 

The performance of the MCMC depends on two issues: the `initialization bias' and the `autocorrelation in equilibrium' (\citealt{sokal96}). The first one refers to the need of reaching the equilibrium distribution and therefore `erasing' any dependence on the initial conditions. This can be achieved by estimating the autocorrelation time in the burn-in phase, and producing chains long enough for them to move away from the initial conditions. The burn-in phase  will then be discarded and not included in the posterior distributions. The second one implies that once the chains have reached equilibrium, we need to take enough samples to obtain a statistically-significant estimate of the errors.  Therefore the performance of MCMC will strongly depend on the length of the chains and the number of walkers, but also on the election of the priors. To ensure a good performance for our analysis we use 300 walkers, a burn-in phase of 500 iterations ($\sim$10 times the autocorrelation time) which are discarded, and 500 iterations after the burn-in phase to sample the posterior distributions.

\begin{figure*}
\begin{center}
\includegraphics[width=15cm]{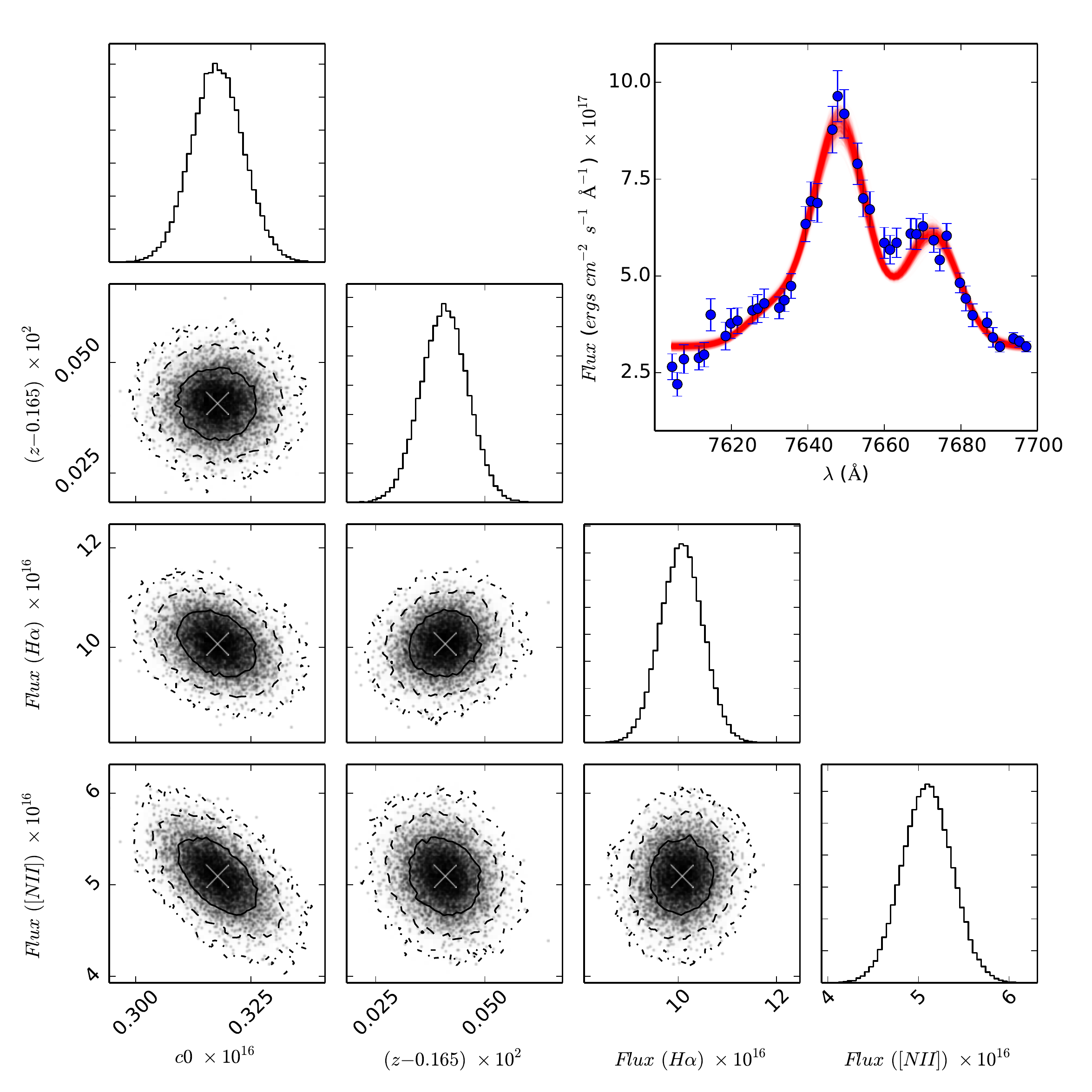}
\caption{Parameter-parameter likelihood distributions with one- (solid), two- (dashed) and three- (dash-dotted) sigma contours obtained for the four-parameter model of the $R_{\rm tot}$ aperture spectrum of galaxy 42713 (the same galaxy shown in Fig.\,\ref{two_spectra}). On the top right we show the observed spectrum (blue points) together with the model fits (red lines). We plot only 1/100 of all the individual MCMC fits for clarity, but they are all very similar. The median values of the parameters' probability distribution are plotted as a white cross on the two-dimensional distributions. The parameters shown are the continuum flux density, the redshift, and the integrated fluxes of the \Ha and \NIIb emission lines. Fluxes are given in \textit{c.g.s} units. }
\label{mcmc}
\end{center}
\end{figure*}

An example of the performance of this algorithm using a four-parameter model (described below) for the spectrum of one of our galaxies is shown in Fig.~\ref{mcmc}. 
Unless otherwise specified we summarise our fit results using the median value of the (marginalised) posterior distribution of each parameter, and estimate uncertainties from the 16th- and 84th-percentiles.

A crucial element of this process is to set the priors that appropriately constrain the regions of parameter space to be explored. From what we already know of our model and data, we set the following priors:
\begin{enumerate}
\item $F_{[\mathrm{N}\textsc{ii}]\,\lambda 6583} > 0$;
\item redshift consistent with the \Ha line being within the wavelength range probed by our spectrum, with an additional 20\,\AA{} window on each side; and
\item $F_{[\mathrm{N}\textsc{ii}]\,\lambda 6583} < 3\,(F_{\mathrm{H}\alpha} + \mathit{continuum})$.
\end{enumerate}
The first of these forces the flux of \NIIb (which, physically, can never be negative) to be positive; the second assumes that the \Ha line lies within the spectral range and the third sets an upper limit for the \NIIb flux such that it cannot be larger than 3 times the flux of \Ha (which is the maximum value of \ratio observed in the WHAN diagram \citealt{cid10}), including the possibility of up to 3\,\AA{} equivalent-width of absorption in \Ha. 

As mentioned above, we define our model using six different parameters.  This number can be reduced if we assume some properties of our spectra. First, one can assume that the slope of the continuum is roughly the same for all the galaxies, which is reasonable considering the short wavelength range covered by the OMEGA spectra ($\leq 200$\;\AA). Secondly, the width of the emission lines will be limited by the instrumental resolution and only substantially broadened in cases of strong AGN (1000 - 25000 km\,s$^{-1}$).
Due to the low spectral resolution and the faintness of some of the objects, adding these constraints will improve the reliability of our estimates.  

\begin{figure}
\begin{center}
\includegraphics[width=.5\textwidth]{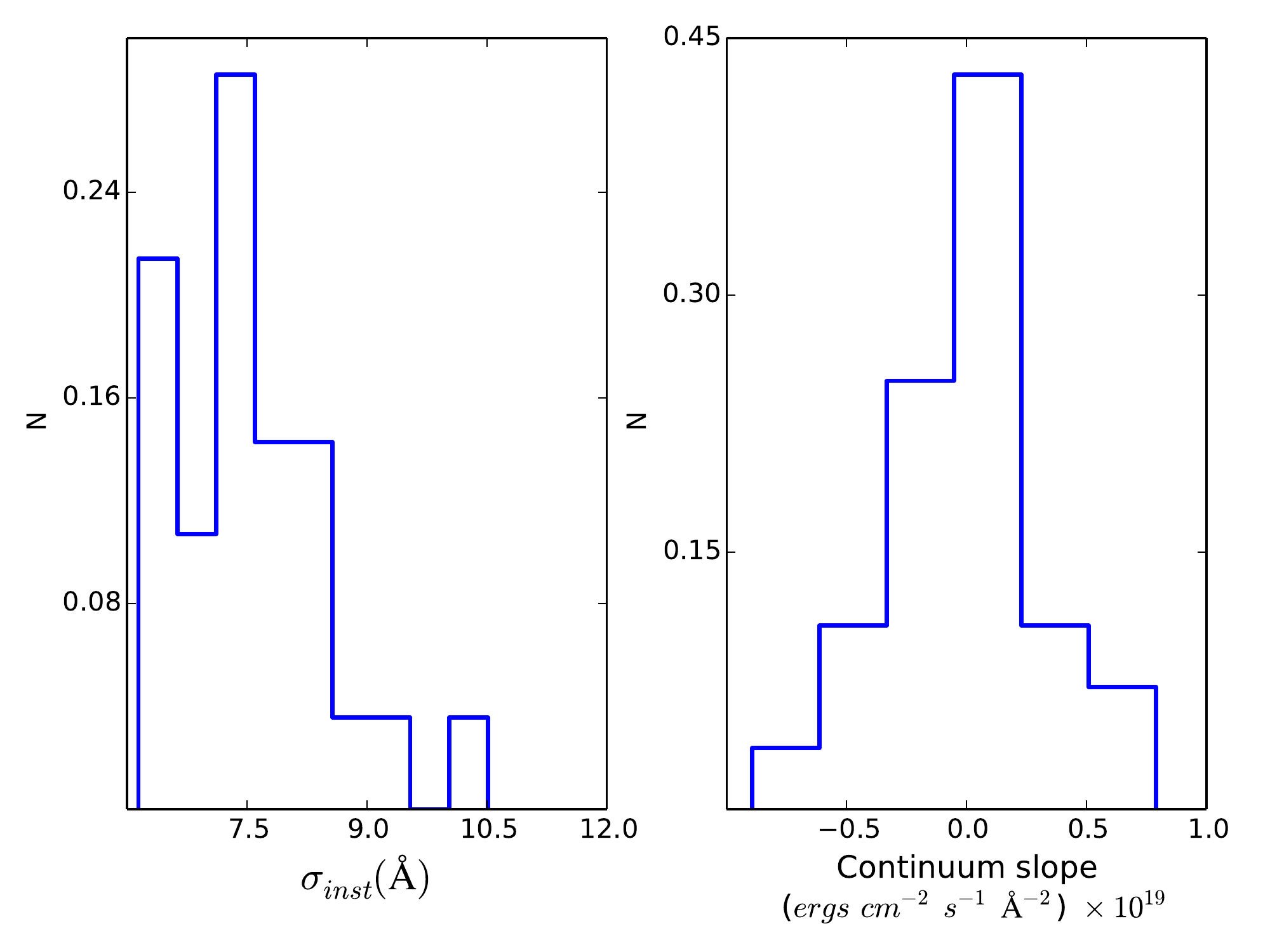}
\caption{Distribution of the width ($\sigma$) of the spectral lines and the continuum slope for a sample of 28 galaxies with very clear emission lines. As a reference, the continuum level of a m$_R$~=~20 galaxy is $\sim5 \times 10^{-18}$ ergs cm$^{-2}$ \AA$^{-1}$. }

\label{histograms}
\end{center}
\end{figure}

In order to find a fixed value for the slope of the continuum and for the instrumental width of the lines, we first performed six-parameter MCMC runs considering only those galaxies that show clear emission lines. We then select galaxies with less than 15 per cent uncertainty in the estimation of the flux and width of the \Ha line, and the median value of the redshift consistent with \Ha and \NIIb being within the wavelength range.  In total we find 28 galaxies that fulfill these requirements.  For these galaxies, Fig.~\ref{histograms} shows the distribution of the line-width and the slope of the continuum.  The continuum slope is consistent with zero, while the distribution of the width of the lines shows a clear peak at 7.5\,\AA. These findings allow us to confidently fix these two parameters and use only four-parameter models to fit the spectra.

\subsection[]{Defining the OMEGA sample of ELGs}
\label{sampledefinition}
After fitting the \Ha and \NII lines as described above, in this section we describe the different criteria we use to define the sample of emission-line galaxies we will use in the remainder of the paper.
We start with all objects in the STAGES catalogue with $16\le~m_R~\le23.0$  that are detected in both COMBO-17 ({\tt combo\_flag} $>$ 1) and in HST ({\tt stages\_flag} $>$ 1). At this stage we do not require them to also be flagged as cluster members according to the definition of \citet{gray09}.\footnote{Membership of the A901/2 cluster structure is determined taking into account the COMBO-17 photometric redshifts and their uncertainties. Given the high accuracy of these redshifts, the procedure maximises completeness without introducing unacceptable contamination. Details ara given in Sect.~3.3 of \citet{gray09}. See, in particular, their Fig.~13 and Equation~8.}

We use measurements made within the total aperture, $R_{\rm tot}$, as defined in the previous section. Our final sample of ELGs in F21 and F22 contains 124 objects fulfilling the following selection criteria:
\begin{enumerate}
\item $\mathrm{relative\,\,error}(\mathrm{W}_{\mathrm{H}\alpha}) < 0.15$;
\item probability of detecting the \Ha line in emission,  $P(F_{\mathrm{H}\alpha}\!>\!0) > 99.7$; and 
\item the \Ha line falls within the wavelength range probed by our observations.
\end{enumerate}
Out of these 124 objects, 103 have been morphologically classified by the STAGES collaboration \citep[see, e.g.,][]{wolf09,maltby10}. 
The galaxies without morphological classification have $z_{\rm phot}>0.4$, the STAGES cut-off for visual classification. Of the 124 ELGs, 72 are classified as clusters members in the STAGES catalogue based on their COMBO-17 photometry ({\tt combo\_flag} $>$ 3).

For the parts of our analysis that make use of both \Ha and \NII measurements (e.g., Sect.~\ref{AGNdiagnostic}), we define a sub-sample requiring both \Ha and \NIIb to be within our wavelength range. This subsample contains 82 ELGs with both $R_{\rm tot}$ and $R_{\rm PSF}$ aperture measurements.

\subsection[]{Detection efficiency and limits}
\label{completeness}

\begin{figure}
\begin{center}
\includegraphics[width=0.5\textwidth]{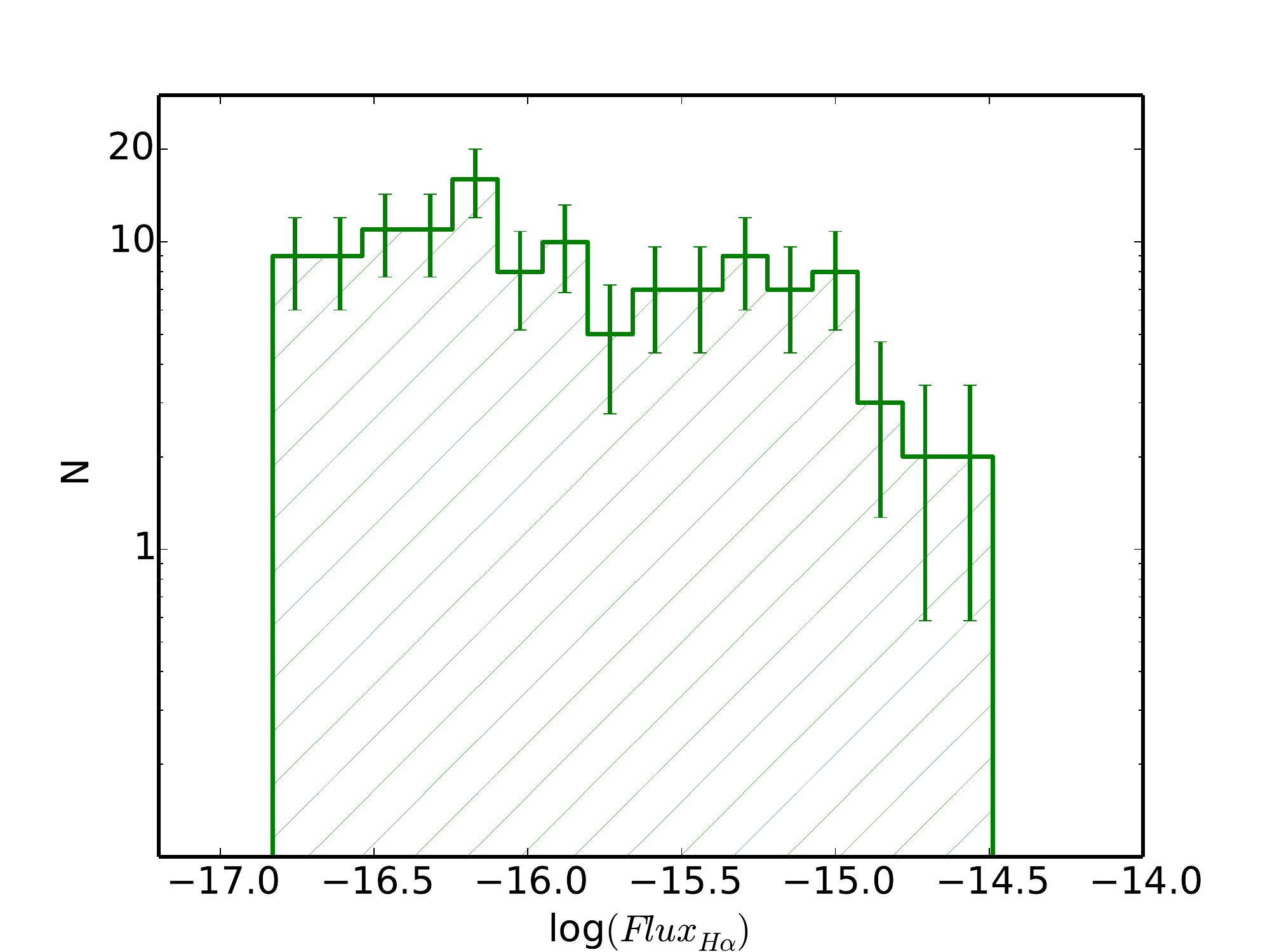}
\caption{The distribution of the \Ha flux for the 124 ELGs in our sample, where the error bars are $1/\sqrt{N}$.}
\label{FHa}
\end{center}
\end{figure}

\begin{figure}
\begin{center}
\includegraphics[width=0.5\textwidth]{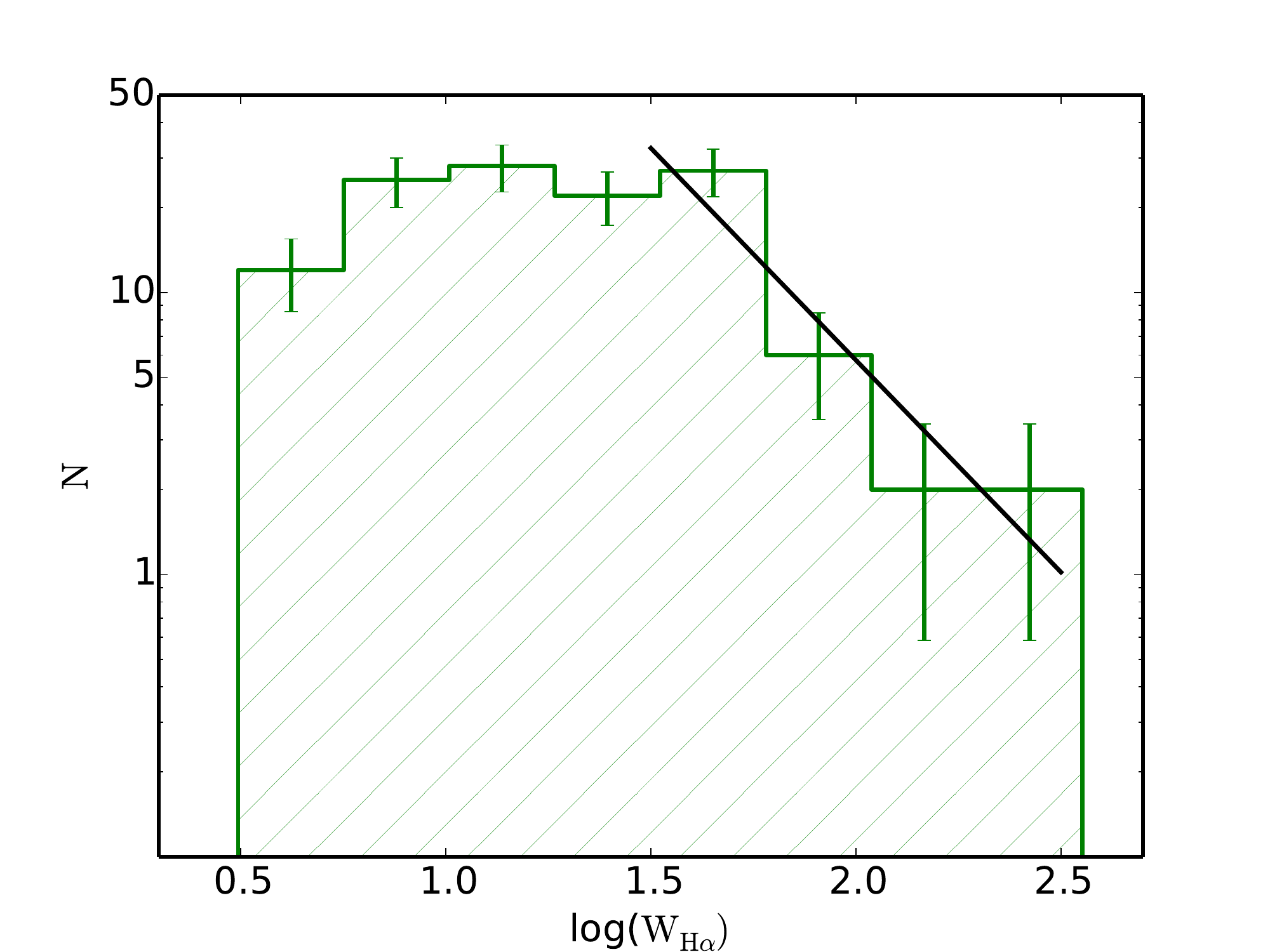}
\caption{The distribution of \WHa for the 124 ELGs in our sample, where the error bars are $1/\sqrt{N}$. 
For equivalent widths larger than $\sim20$--$30$\AA\ the distribution declines steeply with incompleteness kicking in below $\sim5$\AA. We have fitted a line with slope -1.5 and an intercept of 3.8. 
}
\label{WHa}
\end{center}
\end{figure}

Our ability to detect \Ha emission depends on a complex interplay between the brightness of a galaxy, its size, and its \Ha flux and equivalent width. For instance, for a given \Ha flux, galaxies with stronger continuum will be harder to detect. Figs.~\ref{FHa} and~\ref{WHa} show the distribution of the \Ha  fluxes and equivalent widths (\WHa) for the 124 ELGs in our sample. We are able to detect galaxies with \Ha fluxes as low as $\sim1.3\times10^{-17}\,$erg$\,$cm$^{-2}$s$^{-1}$. Given the relatively small sample presented in here (only galaxies in F21 and F22), we leave for a later paper a detailed discussion of the \Ha flux and equivalent width distributions and their dependence on the intrinsic properties of the galaxies and the environment. Reassuringly, the \WHa distribution we obtain is broadly compatible with the one found by \cite{balogh04}. For equivalent widths larger than $\sim20$--$30$\AA\ the distribution declines steeply, following roughly a power law. At lower equivalent widths the distribution flattens out until incompleteness probably kicks in below $\sim5$\AA. When interpreting our data it is important to bear in mind that the sensitivity of our survey declines for lower equivalent widths. 

Due to the complex nature of the \Ha detection process, it would be very difficult to model accurately the detection limits and the completeness of our sample from first principles. Instead, we follow an empirical approach that takes advantage of the wealth of data available from the STAGES database.
\Ha emission is expected to be present in blue, star-forming galaxies. Moreover, many red spirals have been shown to have a certain degree of star formation (\citealt{wolf05}). 
Here we compare our \Ha detections with the numbers of blue-cloud and red spiral galaxies (star forming candidates) in the STAGES catalogue in order to evaluate how efficient the OMEGA survey is at detecting star-forming galaxies and to assess the sample completeness.

\begin{figure}
\begin{center}
\includegraphics[width=0.5\textwidth]{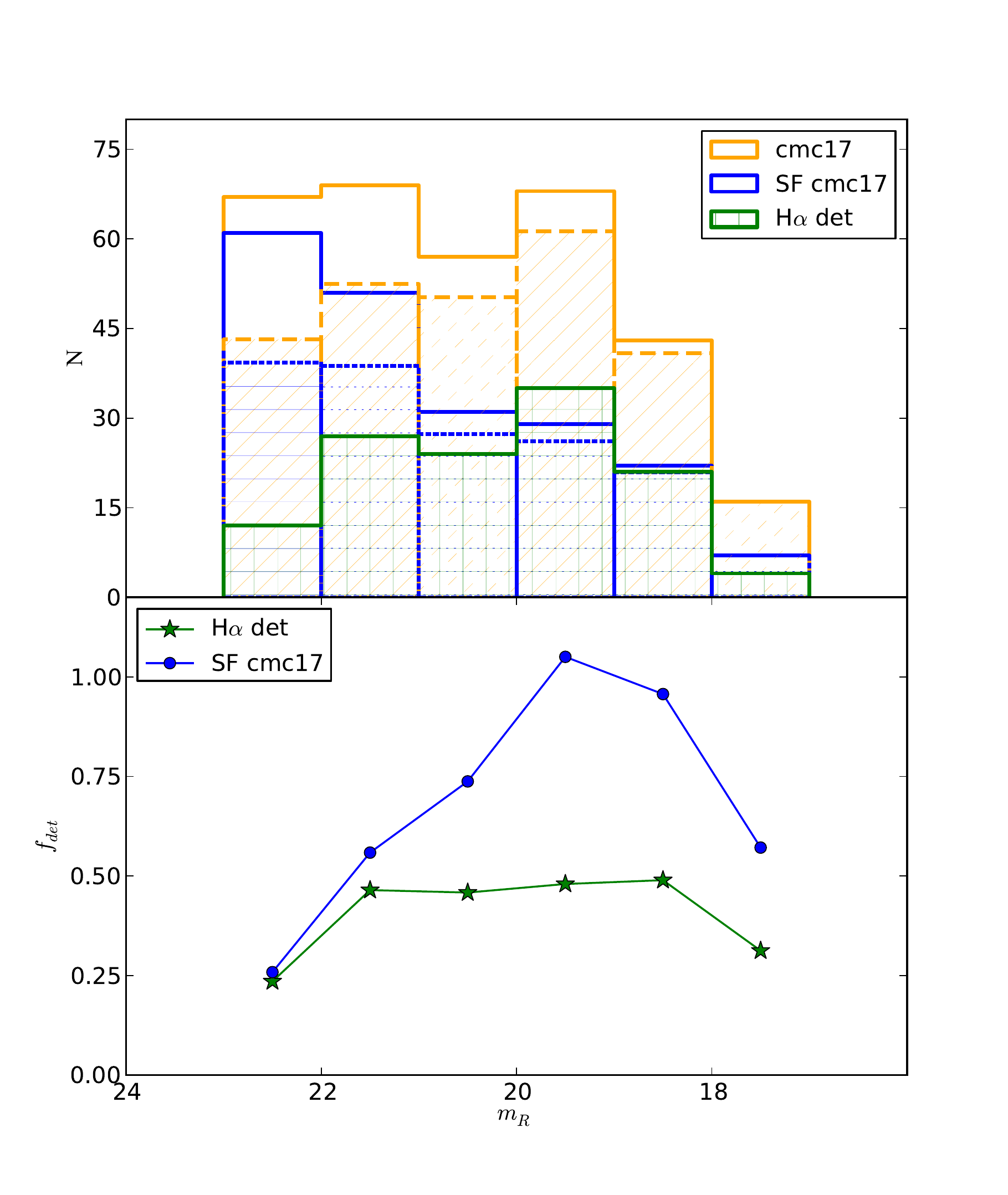}
\caption{\textit{Top panel}: the $m_R$ distribution of galaxies determined to be cluster members using the COMBO-17 photometric redshifts (sample denoted cmc17, in yellow), and the subset of these that are expected to be SF (galaxies classified as blue-cloud or red spiral, in blue). The dashed histograms indicate the same cmc17 and SF cmc17 samples, but corrected for completeness and contamination as discussed in the text. The green square-filled histogram is the distribution of galaxies for which \Ha has been detected in our OSIRIS spectra.
\textit{Bottom panel}: The \Ha detection rate, expressed as a fraction of the full sample of cluster members ($f_{\mathrm{H}\alpha\,\mathrm{det}}$; green stars) and as the fraction of SF cluster members ($f_{\mathrm{SF H}\alpha\,\mathrm{det}}$; blue circles), plotted as a function of $m_R$. See text and equations~\ref{fdet1} and~\ref{fdet2} for details.}
\label{Rmag}
\end{center}
\end{figure}

In the top panel of Fig.~\ref{Rmag} we show the $m_R$ distribution for the galaxies in our sample where \Ha has been detected (green). For comparison, we overplot in yellow the distributions for all COMBO-17 cluster galaxies in the same field-of-view and in blue those identified as star forming (SF). SF galaxies are either blue-cloud or red-spiral galaxies \citep{gray09}. The dashed histograms show the same COMBO-17 samples corrected for completeness and contamination (see below).  
One caveat here is that we do not separate the so-called ``retired'' galaxies \citep{cid11,yb12}, which also have \Ha emission, from the star-forming ones. More details are given in Sect.~\ref{AGNdiagnostic}, when we discuss AGN activity.

The parent sample of COMBO-17 galaxies in F21 and F22 contains $\sim 1400$ objects, with 326 galaxies expected to be cluster members (listed in the STAGES catalogue as having $16.0 \leq m_R \leq 23$ and {\tt combo\_flag} $>3$) with stellar masses $9.0 \leq \log{(M_{\star}/M_{\sun})} \leq 11.5$.
Of these, 203 are either red spirals ({\tt sed\_flag\_cl} $=$ 2) or belong to the blue cloud ({\tt sed\_flag\_cl} $=$ 3), both thus considered to be star-forming galaxies, based on their COMBO-17 SEDs. We securely detect \Ha in 124 objects. From this we estimate that we are able to detect \Ha in  $\sim40$\% of the cluster members and in $\sim60$\% of the cluster star-forming galaxies more massive than $10^{9.0}M_{\sun}$. 

As described in \cite{gray09}, the cluster membership was determined using COMBO-17 photometric redshifts. Despite the relatively high accuracy of these redshifts, the cluster sample suffers from a certain degree of incompleteness and contamination, particularly at faint magnitudes. Therefore, the ``true'' (corrected for incompleteness and contamination) number of cluster galaxies 
$N_\mathrm{cmc17}^\mathrm{c}$ can be estimated as
\begin{equation}
\label{corr1}
N_{\rm cmc17}^{\rm c}=\frac{N_{\rm cmc17}(1-\mathit{cont}_\mathrm{frac})}{\mathit{comp}_\mathrm{frac}},
\end{equation}
where $N_{\rm cmc17}$ is the observed number of cluster members based on their photometric redshifts. In this equation, $\mathit{comp}_\mathrm{frac}$ and $\mathit{cont}_\mathrm{frac}$ are the magnitude-dependent completeness and contamination fractions respectively (see Fig.~14 of \citealt{gray09}). Similarly,  the ``true'' number of SF cluster galaxies 
$N_\mathrm{SFcmc17}^\mathrm{c}$ is
\begin{equation}
\label{corr2}
N_{\rm SFcmc17}^{\rm c}=\frac{N_{\rm SFcmc17}(1-\mathit{cont}_\mathrm{frac})}{\mathit{comp}_\mathrm{frac}},
\end{equation}
where $N_{\rm SFcmc17}$ is the observed number of SF cluster galaxies.

The bottom panel of Fig.~\ref{Rmag} shows as green stars the fraction of cluster galaxies detected in \Ha, defined as 
\begin{equation}
\label{fdet1}
f_{\mathrm{H}\alpha\,\mathrm{det}}=\frac{N_{\mathrm{H}\alpha\,\mathrm{det}}} {N_{\rm cmc17}^{\rm c}}, 
\end{equation}
and as blue circles the fraction of SF cluster galaxies detected in \Ha
\begin{equation}
\label{fdet2}
f_{\mathrm{SF H}\alpha\,\mathrm{det}}=\frac{N_{\mathrm{H}\alpha\,\mathrm{det}}} {N_{\rm SFcmc17}^{\rm c}}.
\end{equation}

We effectively detect \Ha for all SF cluster members brighter than $m_R \sim 20$. Fainter than this, the completeness drops off, as expected -- only galaxies that are bright in \Ha are detected. Note that the apparent drop for galaxies brighter than $m_r \sim 19$ is due to low-number statistics in the brightest bins. 
As a fraction of all cluster-members, the completeness does not depend strongly on $m_R$ since at lower masses a greater fraction of cluster galaxies are star forming.  
Note that $f_\mathrm{det} > 1$ means that we are detecting \Ha in more galaxies than are expected to be star forming (based on the {\tt sed\_flag\_cl} flag). This reflects both the uncertainty in the COMBO-17 SED classification and the high sensitivity of our survey. 

\section[]{Results}
As a preview of the kind of scientific issues that we will be able to address using OMEGA data, in this section we identify star-forming galaxies and AGN from our ELG sample, and investigate their properties, including mass and morphology. This is done by cross-matching our OMEGA observations with the STAGES catalogue (\citealt{gray09}). In addition, we present the \Ha luminosity function (LF) in the densest regions in order to understand how comparable our \Ha SFRs are to literature values of other sources. We finally compare our \Ha-derived star-formation rates (SFRs) to UV and IR ones from \cite{gray09}. In subsequent papers we will present a full analysis for the complete dataset.

\subsection{AGN and star-forming galaxies}
\label{AGNdiagnostic}
\begin{figure}
\begin{center}
\includegraphics[width=.52\textwidth]{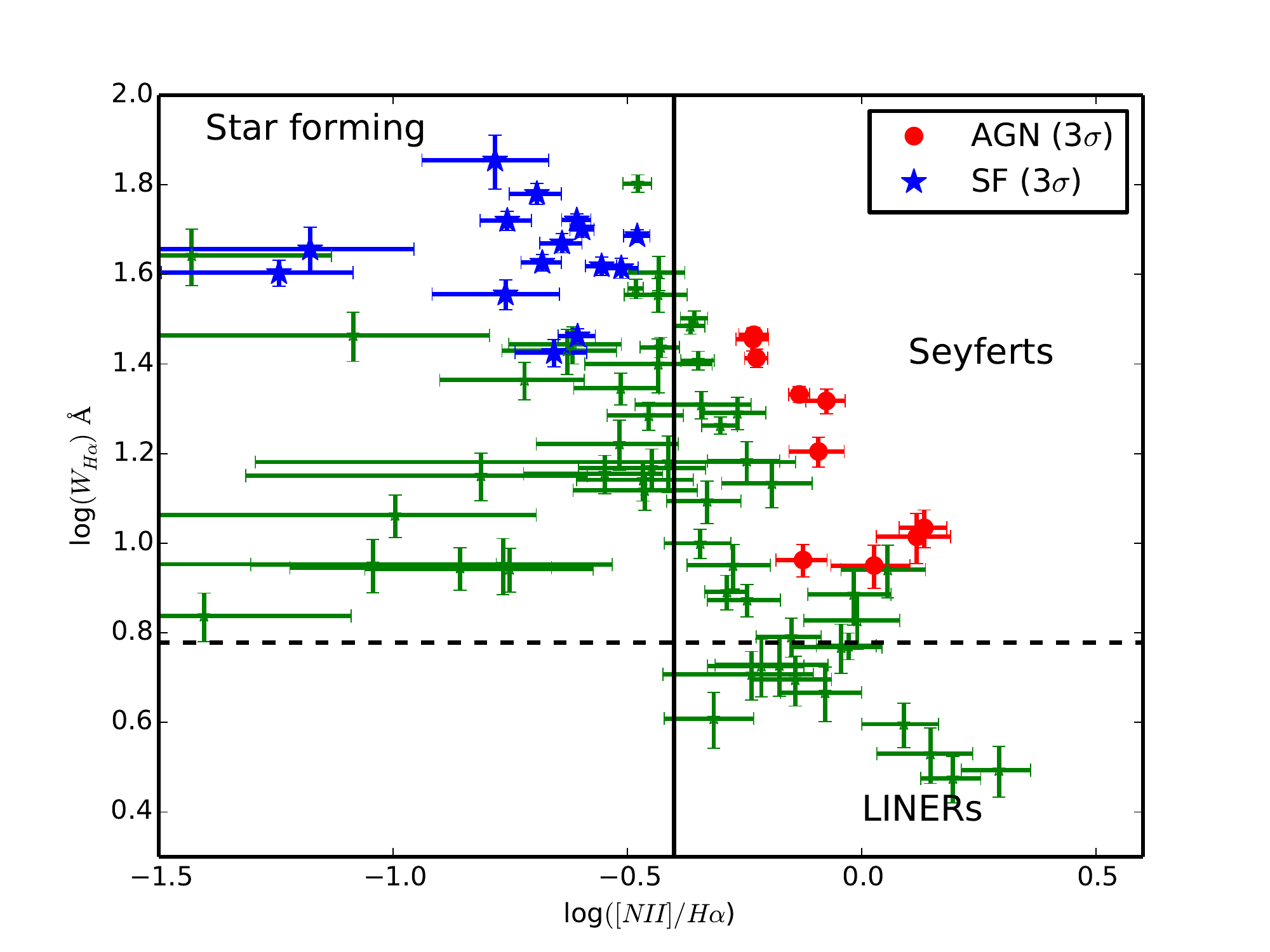}
\caption{A diagnostic plot of \ratio vs. \WHa, the so-called WHAN diagram (\citealt{cid10}). The vertical and horizontal lines separate regions with different sources of ionisation \citep{cid10,cid11}. We only show detections using $R_{PSF}$ aperture measurements where both \Ha and \NII fall in the OMEGA probed wavelength range: a total of 82 ELGs. From these we find 10 secure ($\ge3\sigma$ confidence) AGN hosts (Seyferts; red circles) and 16 star-forming galaxies (SF; blue stars). The green points represent galaxies for which the classification is uncertain and/or contain mixed sources of ionisation.}
\label{WHAN}
\end{center}
\end{figure}

In Fig.~\ref{WHAN} we show the \ratio vs.\@ \WHa (WHAN; \citealt{cid10}, \citealt{cid11}) diagram for our ELGs.
Here we use the $R_{\rm PSF}$ aperture, as this provides the cleanest measurement of the nuclear emission of the galaxy.
The WHAN diagram is an alternative economical form of the widely-used BPT diagnostic diagram (\citealt{bpt}). \cite{cid10} and \cite{cid11} have convincingly shown that the ratio of \NII to \Ha is truly powerful in separating AGN from SF galaxies, especially when the equivalent width of \Ha is also considered.

Using the probability density distributions obtained for the continuum level and fluxes in \Ha and \NII, we can obtain the probability density distributions of the \WHa and the \ratio flux-ratio. Assuming standard values to differentiate sources of ionization in the WHAN diagram (cf. Fig.~\ref{WHAN}), we can therefore calculate the probability for each galaxy in our sample being dominated by star-formation or AGN activity.  Following this, 10 objects have a greater than 99.7 per cent probability ($\ge3\sigma$ confidence) of hosting a dominant AGN (Seyfert; red circles), while 16 have a similar probability of being SF-dominated (blue stars).
The remaining 56 galaxies (green points) have less certain classifications due to the observational uncertainties on  \ratio. Obviously, it is possible to classify galaxies into SF and AGN with less certainty by lowering the probability threshold. This would increase the size of the SF and AGN samples, but the confidence for each individual object would be reduced. Nevertheless, since we have robust probability distributions for the line ratios, the statistical properties of such samples can still be analysed rigorously. It is important to bear in mind, however, that in reality there is no hard division between SF- and AGN-dominated systems. Better observational errors would place the galaxies more accurately on the WHAN diagram, but since the boundaries are somewhat fuzzy (and depend on other physical parameters such as metallicity and inter-stellar medium density), the classifications would, in many cases, remain uncertain: a significant fraction of the green points may well contain mixed sources of ionisation.

One interesting population revealed by the WHAN diagram is the so-called ``retired'' galaxy population (\citealt{cid11}, \citealt{yb12}, \citealt{s15}). They are located in the bottom-right part of such diagram, which is also occupied by LINER-type AGN. These retired galaxies have stopped forming stars, their gas is believed to be ionized by hot low-mass evolved stars, and they do not have a detectable contribution from an AGN. Furthermore, their \Ha emission is often extended \citep{belfiore14}. This population will be addressed in detail in a subsequent manuscript, when we analyse the resolved emission-line images.

In the following we explore how the properties of the SF galaxies and AGN hosts that we have identified compare to one another and to the overall population of ELGs. We consider here their total $m_R$, morphologies and stellar masses.

\begin{figure}
\begin{center}
\includegraphics[width=.5\textwidth]{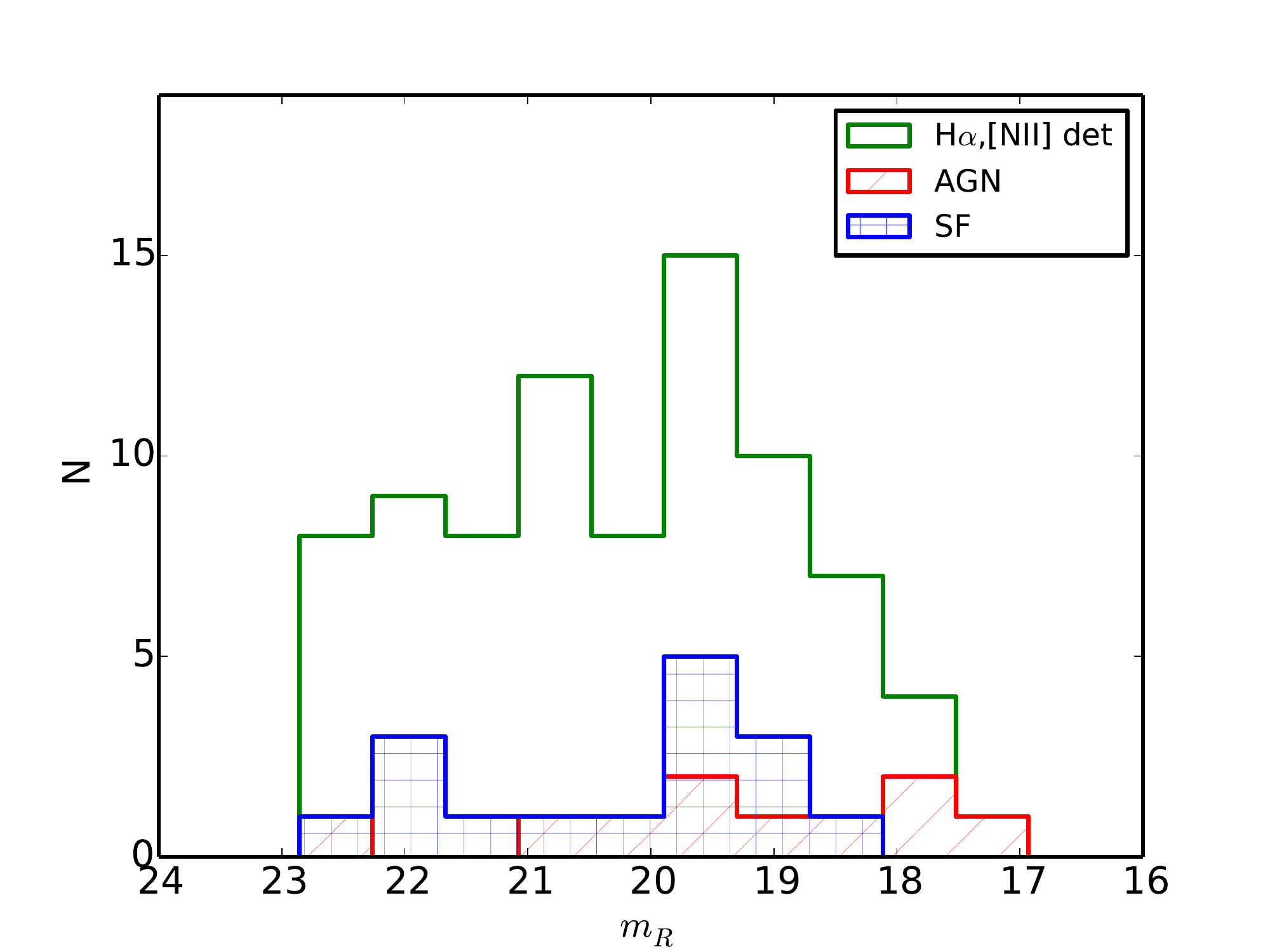}
\caption{The distribution of the $R$-band magnitudes from COMBO-17 for ELGs with detected \NIIb and \Ha (green line), as well as for the subsamples of high-confidence AGN (Seyferts; red histogram) and SF objects (in blue). See text for details.}
\label{Rmag_AGN_SF}
\end{center}
\end{figure}

\begin{figure}
\begin{center}
\includegraphics[width=.5\textwidth]{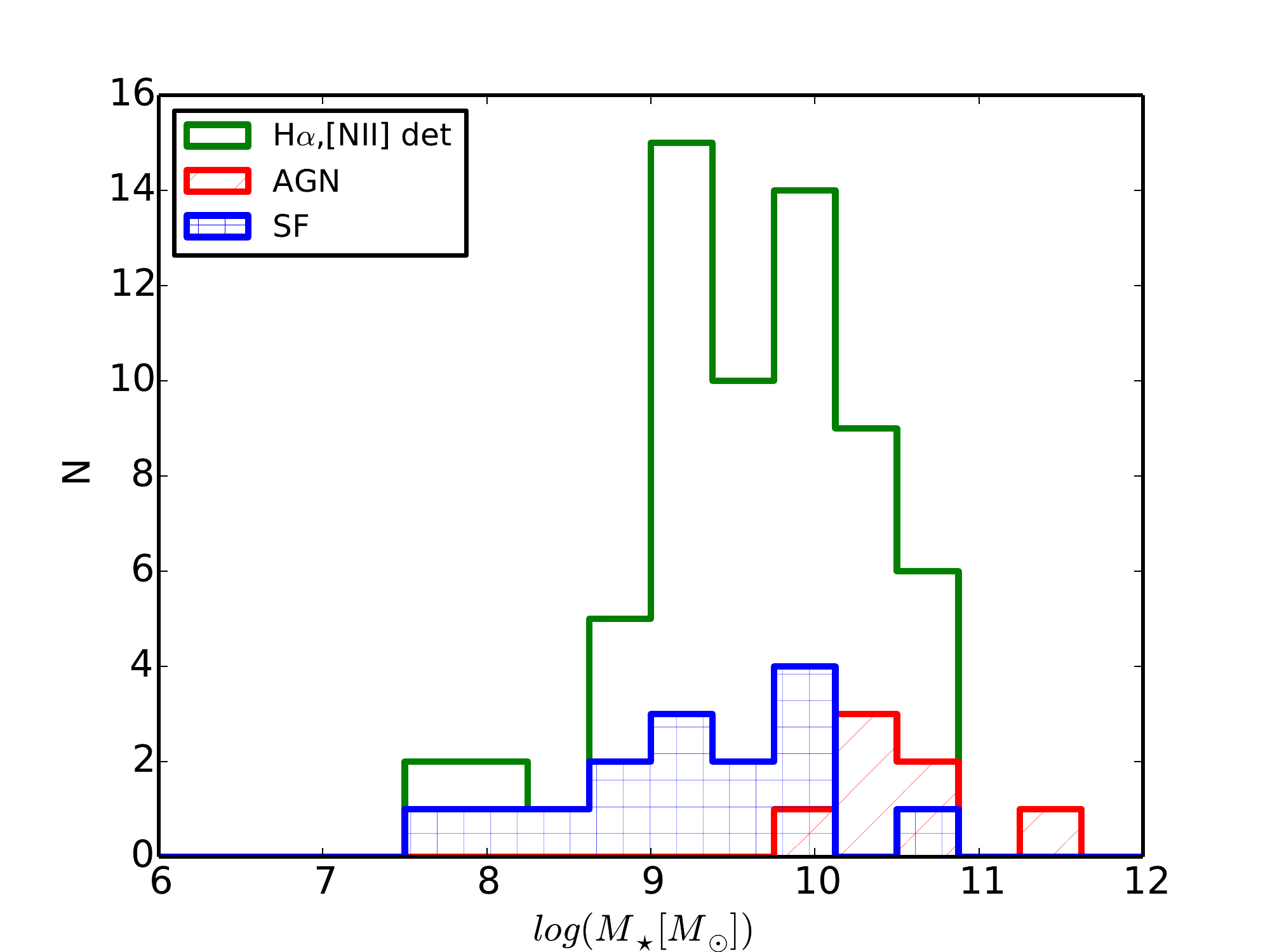}
\caption{Stellar mass distribution for ELGs with detected \NIIb and \Ha, together with the AGN and SF subsamples. Object selection and lines as in Fig.~\ref{Rmag_AGN_SF}.}
\label{logmass_AGN_SF}
\end{center}
\end{figure}

In Figs.~\ref{Rmag_AGN_SF} and \ref{logmass_AGN_SF} we present histograms of $m_R$ and stellar masses for the ELGs with both \Ha and \NIIb secure detections. The distributions of the secure AGN (Seyferts) and SF objects are also shown.
Galaxies hosting AGN are brighter in the $R$-band and more massive than SF galaxies. Moreover, the overall behaviour of the \Ha-detected galaxies appears to mimic the mass and $m_R$ distribution of SF galaxies, suggesting that the majority of the the galaxies with uncertain classifications are SF-dominated.

\begin{figure}
\begin{center}
\includegraphics[width=9cm]{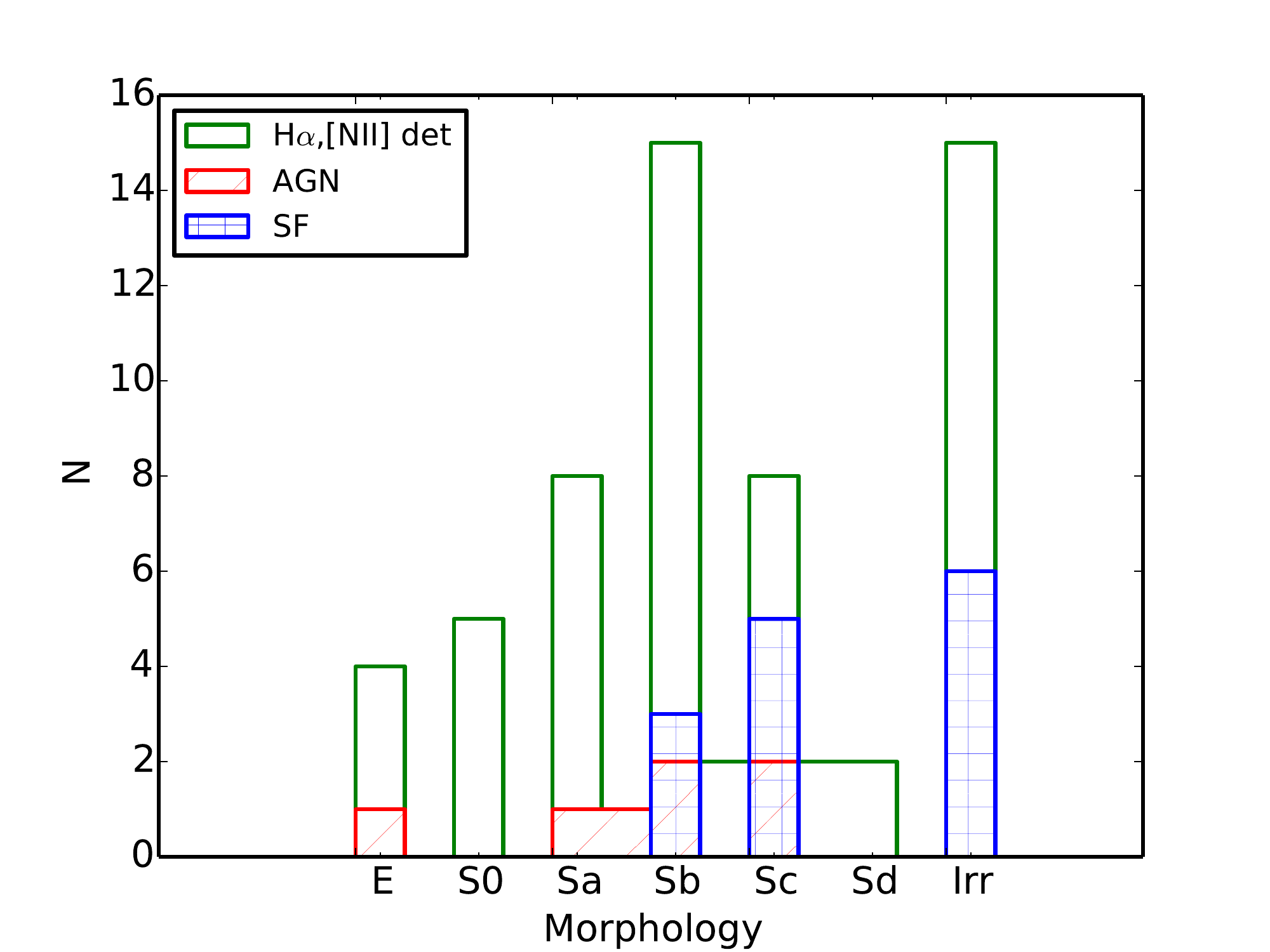}
\caption{The distribution of morphologies for ELGs with detected \NIIb\, and \Ha as well as the subsamples of the high-confidence ($\ge3\sigma$) AGN and SF objects.}
\label{T_SF_AGN}
\end{center}
\end{figure}

In Fig.~\ref{T_SF_AGN} we show the distribution of morphologies.
Galaxies determined to be hosting AGN with $\ge3\sigma$ confidence are typically of earlier-type than those for which the dominant source of ionisation is SF or uncertain.

\begin{figure*}
\begin{center}
\includegraphics[width=15cm]{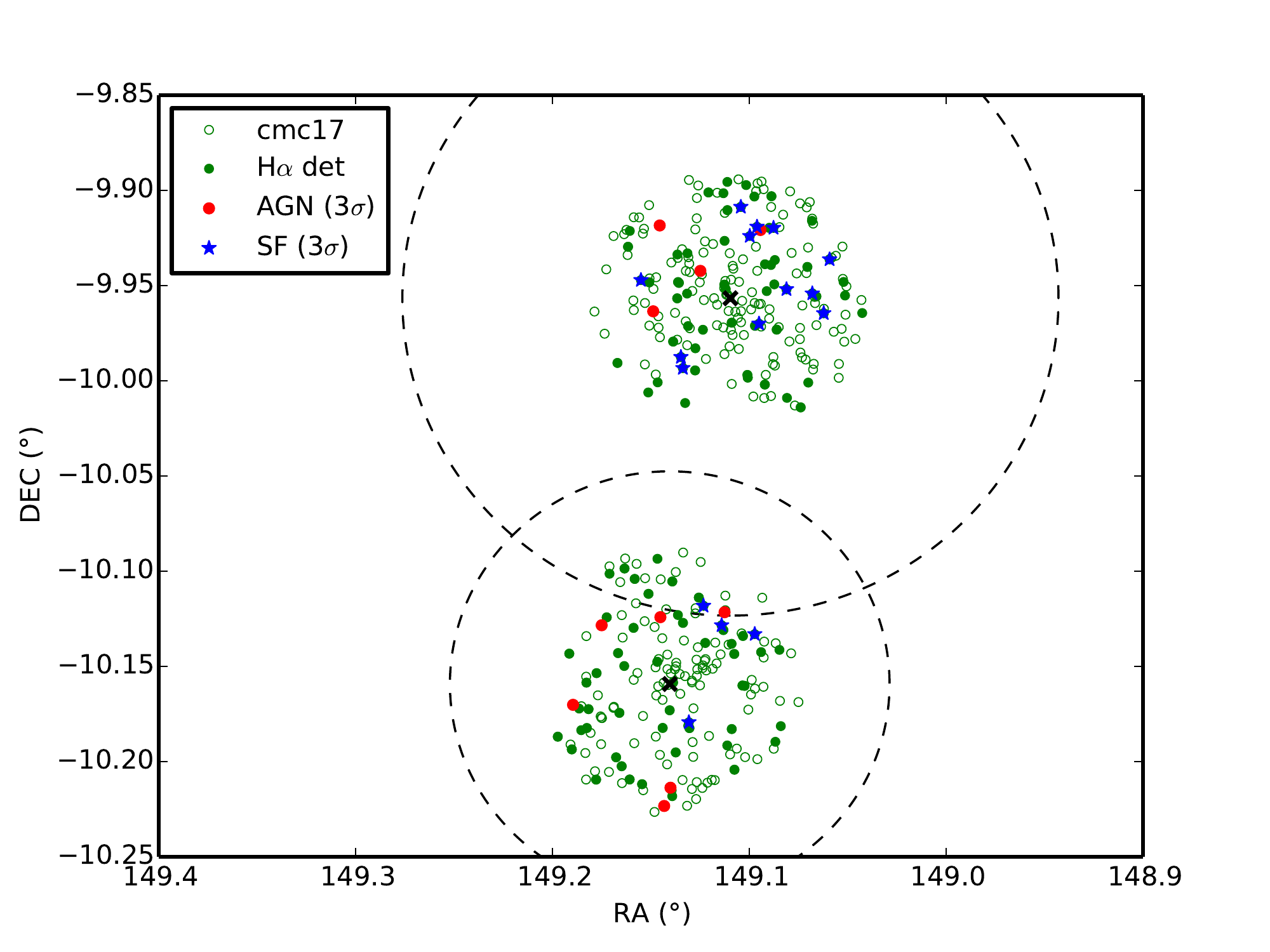}
\caption{The distribution on the sky of the cluster galaxies (as determined by COMBO-17; denoted cmc17 in the legend) in the F21 and F22 regions down to $m_R<23$. ELGs (where \Ha has been securely detected) are represented by green filled circles. Galaxies classified as AGN or star-forming with $\ge3\sigma$ confidence are shown as solid red circles and blue stars respectively. The centres of A901a and A902 (determined from the brightest cluster galaxies) are marked with black crosses. The large dashed circles show the $r_{200}$ radii of A901 and A902 derived from the weak lensing analysis of \citet{hey08}.}
\label{radec}
\end{center}
\end{figure*}

In Fig.~\ref{radec} we show the distribution on the sky of the galaxies in the F21 and F22 regions down to $m_R<23$ that are classified as cluster members in the STAGES catalogue. Galaxies with more than $3\sigma$ confidence of being AGN- or SF-dominated (see Fig.~\ref{WHAN}) are denoted by red and blue filled circles respectively, whereas other \Ha secure detections are indicated by filled green circles. The centres of A901a and A902 are marked with black crosses and are defined as the location of the brightest cluster galaxies.
We note that secure SF and AGN galaxies are located towards the outer parts of the clusters and are completely absent in the central regions, especially when looking at F22.

\cite{gilmour07} presented a catalogue of \textit{XMM}-detected point sources in the A901/902 field (see their table~12). From the 12 X-ray sources that they identify as likely AGN within the cluster, 4 objects fall in the FOV of F21 and F22. Only one of them is part of our ELG sample with securely-detected \Ha (ID11827). However, this galaxy is not classed as an AGN because the \NIIb line used in the classification process falls outside the wavelength range probed by the F22 observations. Reassuringly, the \NIIa line is detected and, based on its intensity, this object has a 93.5\% probability of being an AGN. Moreover, preliminary analysis of an overlapping field (which samples a slightly different wavelength range) indicates that \NIIb is clearly detected for this object, indicating that it is probably an AGN. For a second X-ray AGN (ID41435) neither \Ha nor \NII fall in the wavelength range probed by our observations. 
For the remaining two X-ray AGNs (ID12953 and ID44351) our observations sample both \Ha and \NII but their \Ha lines are below our detection probability threshold ($P(F_{\mathrm{H}\alpha}\!>\!0) < 99.7$; 
see section~\ref{sampledefinition}) and therefore do not make it into our ELG sample. In summary, one out of the four X-ray AGN that we could have detected is also highly likely to be an optical AGN. The others have either too faint emission lines or do not have \Ha or \NII in our probed wavelength range. It is possible that these objects may be heavily obscured Compton-thick AGN and therefore would not be expected to show optical signs of nuclear activity. 
On the other hand, the fact that none of the 10 highly likely optical AGN found in the OMEGA survey are associated with a \cite{gilmour07} X-ray source clearly shows that OMEGA is able to detect a large number of optical AGN with relatively weak or no X-ray emission.

\subsection{The \Ha Luminosity Function}
 
The \Ha luminosity function of star forming galaxies is an important measurement for comparison among different studies and environmental dependencies.
The luminosity of the \Ha nebular emission line is a SFR indicator with a direct physical connection to short-lived, massive stars. In order to calculate the luminosity function of \Ha, line fluxes are converted to luminosities via the following:
\begin{equation}
L_{\mathrm{H}\alpha}=4\pi \, D^{2}_\mathrm{L} \, F_{\mathrm{H}\alpha} \;,
\end{equation}
where $D_\mathrm{L}$ is the luminosity distance (795\,Mpc at $z=0.167$).

\begin{figure}
\begin{center}
\includegraphics[width=.48\textwidth]{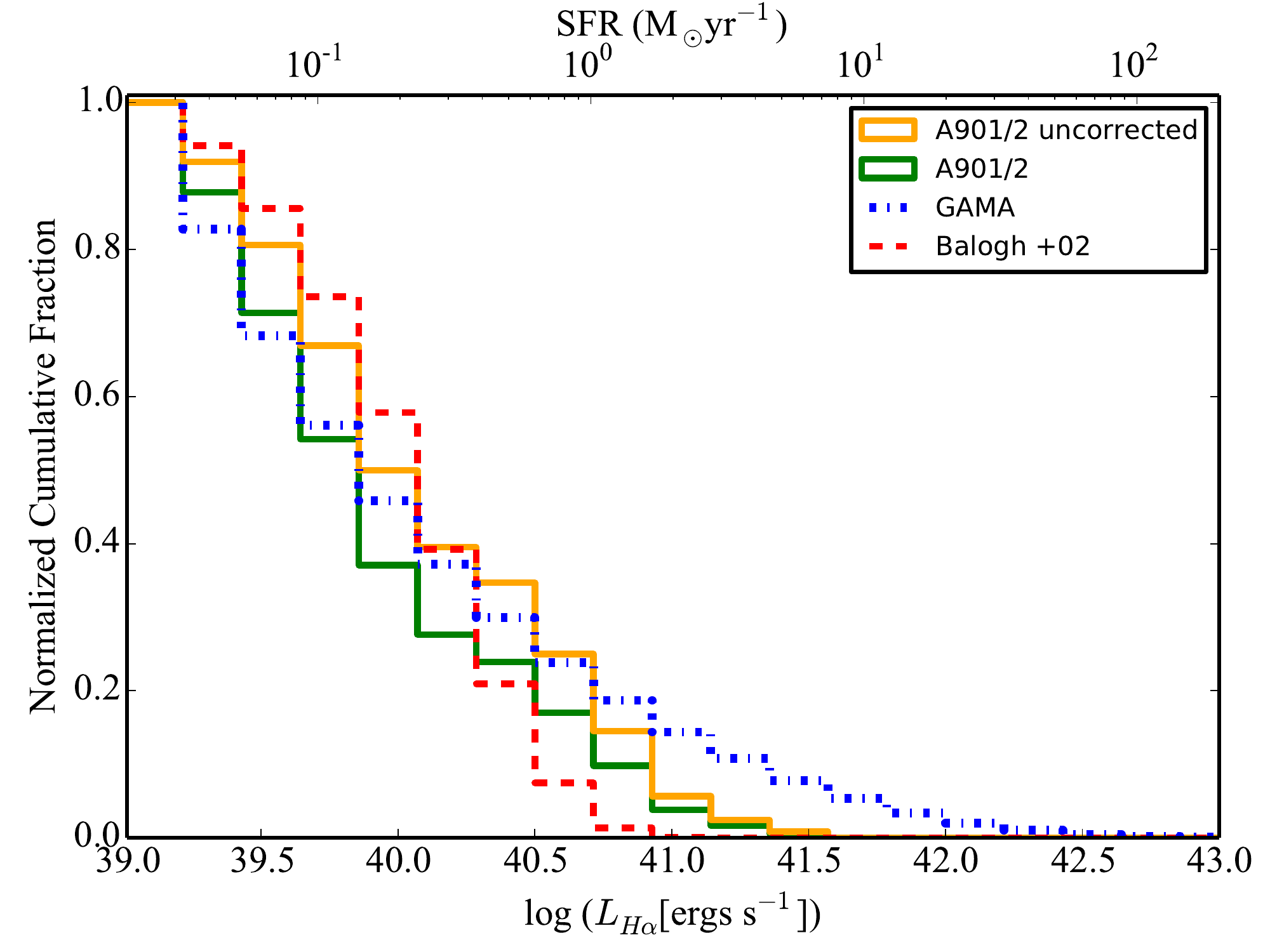}
\caption{The cumulative \Ha luminosity function for the highest density regions of A901/2 corrected (solid green line) and uncorrected (solid yellow line) for contamination and completeness. For comparison,  we show the \Ha LF from the cluster A1689 at $z=0.18$ (blue dash-dot line; \citealt{balogh02}) and the field from the GAMA survey at $z=0.1$--$0.2$ (red dashed line; \citealt{guna13}). The densest regions studied by OMEGA have SFRs that fall in between those of the cluster A1689 and the field. }
\label{lumfunc}
\end{center}
\end{figure}

In Fig.~\ref{lumfunc} we show the \Ha cumulative luminosity function (LF) for the highest density regions of A901/2, using our \Ha detections from F21 and F22 as described in the previous sections. The top axis shows the SFR, assuming 1~mag average extinction, according to:
\begin{equation}
\label{SFReq}
\mathrm{SFR}_{\mathrm{H}\alpha}\,\left[M_{\sun}\,\mathrm{yr}^{-1}\right] = 2 \times 10^{-41} L_{\mathrm{H}\alpha}\,\left[\mathrm{erg\,s}^{-1}\right]\;,
\end{equation}
following \cite{ken98}.

We use the results of Sect.~\ref{completeness} in order to correct the OMEGA \Ha LF for incompleteness. We parametrise the dependence of $f_{\mathrm{SF H}\alpha\,\mathrm{det}}$ on the $R$-band magnitude $m_R$ as
\begin{equation}
f_{\mathrm{SF H}\alpha\,\mathrm{det}}= \left\{ \begin{array}{rl}
 -0.24m_R+5.7 &\mbox{ if $m_R \ge 20$} \\
  1.0 &\mbox{ if $m_R < 20$}
       \end{array} \right.
\end{equation}
We consider our sample to be essentially complete for $m_R<20$ (see bottom panel of Fig.\ref{Rmag}). To correct for incompleteness, we weight the OMEGA \Ha LF by $1/f_{\mathrm{SF H}\alpha\,\mathrm{det}}$.

For comparison we show the A1689 luminosity function (at $z=0.18$) from \cite{balogh02} as well as the field \Ha LF from GAMA (\citealt{guna13}). For the field \Ha LF we use the best-fitting parameters of the Saunders function (\citealt{saunders90}) that \cite{guna13} obtain for the redshift bin $z = 0.1$--$0.2$ ($\log{L^{*}} = 34.55$, $\log{C} = -2.67$, $\alpha = -1.35$, $\sigma = 0.47$) to reconstruct the corresponding normalised cumulative fraction.
For A1689 we use the Schechter function parameters: $\alpha=-0.1$ and $L^{*}=10^{40}\;\mathrm{erg\,s}^{-1}$ given by \cite{balogh02}, to reconstruct the cumulative \Ha LF.
Compared to the field sample from \cite{guna13}, the densest parts of A901/2 contain galaxies with lower, on average, SFRs.
Conversely, A901/2 hosts brighter \Ha
emitters than A1689.  
This is not a surprise, as A1689 is a more evolved and denser cluster than the densest regions of A901/2 (\citealt{lemze09}, \citealt{alamo13}). We conclude that in the densest regions of the A901/2 system star formation has been significantly suppressed compared to the field, but not as much as in denser, more evolved, cluster environments.

\subsection{Star Formation Rates}

The SFR-stellar mass relation is a useful diagnostic for the star-formation properties and history of a galaxy population. In this parameter space there is a reasonably well-defined main sequence whose properties have been characterised at different redshifts (see e.g. \citealt{whitaker12}).
Several tracers across the electromagnetic spectrum may be used to estimate the star formation rate of a given galaxy. The STAGES catalogue of \cite{gray09} provides UV and IR SFRs for the A901/2 region.
Rest-frame UV light originates mainly from massive stars and thus directly traces young stellar populations (with main sequence lifetimes $\la\,10^8$\,yrs); however, it is strongly attenuated by dust. The SFR from the UV continuum is often dubbed `current SFR'. 
The interstellar medium associated with a star-forming region can be quite dusty, so a significant fraction of the UV luminosity produced by massive young stars can be absorbed. This radiation heats the dust and is re-emitted in the far infrared.
Moreover, the interstellar medium around young massive stars is ionised by the Lyman-continuum photons they emit. The recombination of this ionised gas produces \Ha emission-lines. The SFR from the \Ha nebular emission is often dubbed `instantaneous SFR', tracing stars with main sequence lifetimes of $\la\,2 \times 10^7$\,yrs (\citealt{mo11}).
In the following figures, we adopt an \Ha SFR with 1\,mag of extinction (see Eq.\,\ref{SFReq}).
At this point, we do not attempt to subtract the AGN contribution to the SFR from the \Ha fluxes. Resolved star formation properties will be studied in a subsequent manuscript.

\begin{figure}
\includegraphics[width=.47\textwidth]{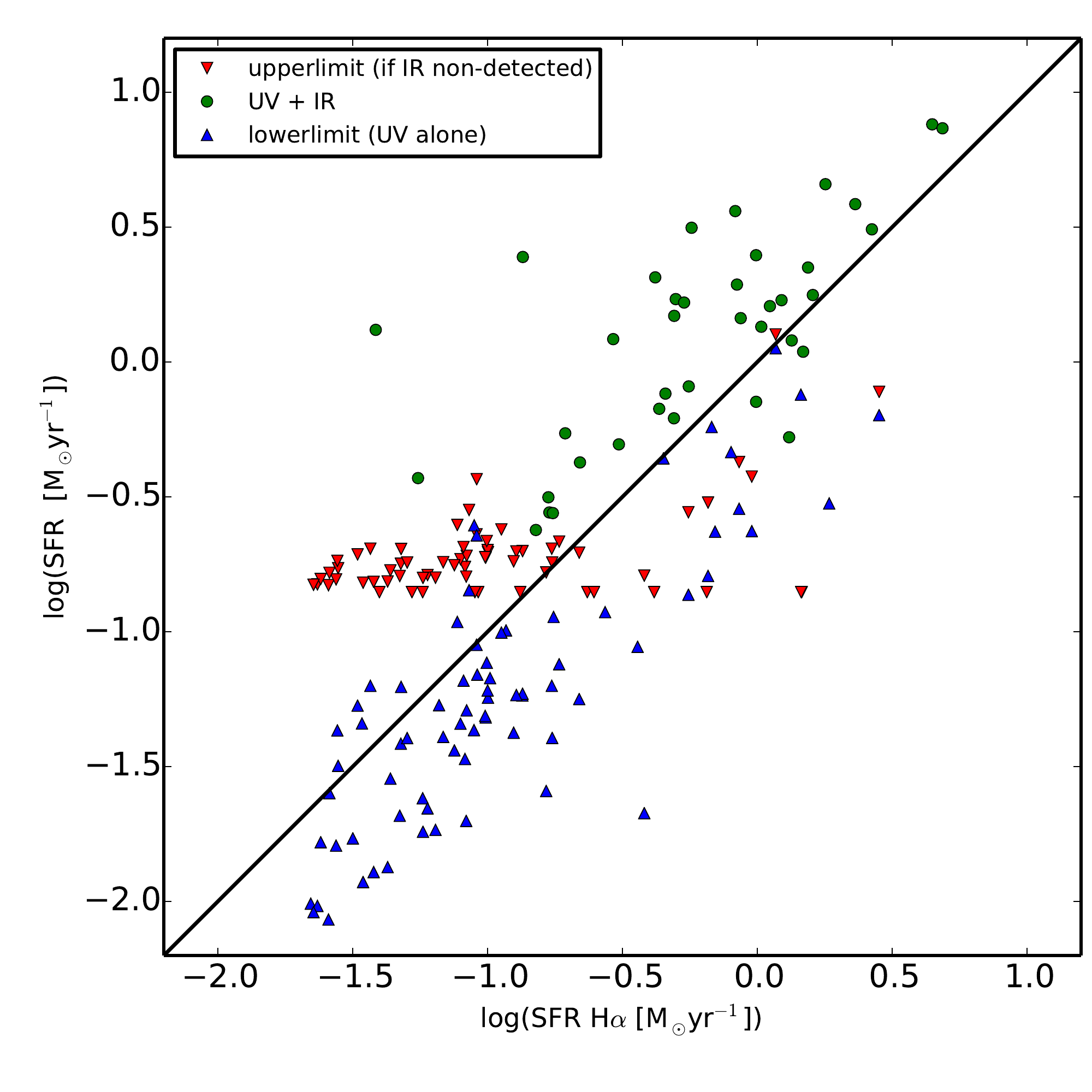}
\caption{The $\log{\mathrm{SFR}_{\mathrm{H}\alpha}}$ determined from $R_\mathrm{tot}$ aperture measurements compared with other estimates of $\log{\mathrm{SFR}}$. See legend and text for details on the meaning of the different datapoints. 
}
\label{SFRHa_UV_IR}
\end{figure}

We compare our \Ha SFR against those from other available tracers in Fig.~\ref{SFRHa_UV_IR}.
The UV SFRs are based on the UV luminosity estimated using the 2800\AA\ rest-frame luminosity from COMBO-17, which has a complete coverage around the A901/2 multi-cluster system. 
The UV SFRs have been corrected for extinction as part of the SED-fitting procedure. Details can be found in \cite{gallazzi09} and \cite{wolf09}.
The IR SFRs are determined from Spitzer/MIPS 24$\micron$ data. In the cases where there was no detection in the IR, upper limits to the SFRs were estimated (\citealt{wolf09}).
The green datapoints (denoted `UV+IR' in the figure legend) represent the composite SFR measurements.
Note that these points, which account for the obscured and unobscured star formation, are, on average, $\sim 0.3$\,dex above the one-to-one line. This difference may be due to the \Ha SFR estimates missing heavily-obscured SF, but it could also be indicative of a declining SFR, given the different timescales of the indicators.
In the case of the SFRs from the UV alone, the correlation with the \Ha SFR is stronger, which is not surprising as the UV and \Ha are sensitive to comparable SF timescales, with the UV tracing slightly longer ones than \Ha. However, there is also a clear $\sim0.3$\,dex offset since the effect of extinction is significantly stronger in the UV than in the red.

\begin{figure}
\includegraphics[width=0.515\textwidth]{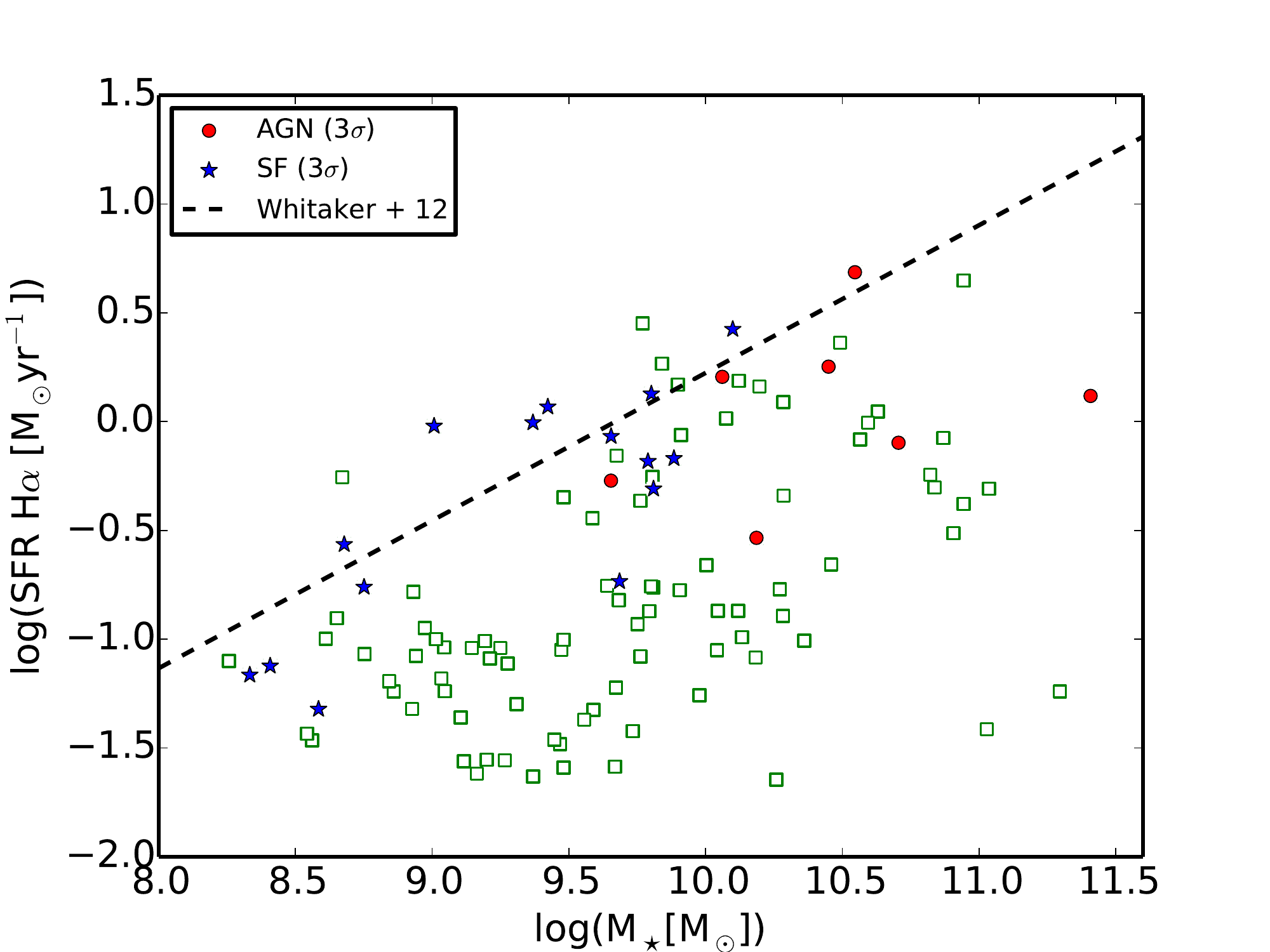}
\caption{Stellar mass vs. SFR(\Ha) determined from $R_\mathrm{tot}$ aperture measurements. The galaxies with a 3$\sigma$ confidence of being AGN- or SF-dominated, with selection based on the PSF aperture (see Fig~\ref{WHAN}) are denoted by red circles and blue stars respectively. The dashed-line is a SFR-mass relation for field star-forming galaxies at $z=0.167$ (\citealt*{whitaker12}). Comparing our secure \Ha detections to the dashed-line representing the field galaxies indicates a suppression of star formation in these high density regions.
}
\label{Mass_SFR}
\end{figure}

In Fig.~\ref{Mass_SFR} we show the stellar mass vs. the \Ha-determined SFR. Galaxies reliably classified as AGN and SF objects (cf. section~\ref{AGNdiagnostic} and Fig.~\ref{WHAN}) are indicated by red circles and blue stars respectively. 
At a given stellar mass, only the galaxies that are brightest in \Ha can be reliably classified as SF or AGN, while the rest (green squares) have either mixed ionisation sources or uncertain classification, but they are expected to be dominated by star formation (section~\ref{AGNdiagnostic}). 
For comparison, the solid line shows the parametrized SFR-mass relation derived from equations~1, 2 and 3 of \cite{whitaker12} for star-forming field galaxies at the redshift of A901/2,
\begin{equation}
\label{sfr_mass}
\log{\mathrm{SFR}} = 0.68\,\log{M_{\star}} - 6.56.
\end{equation}

The stellar masses used by Whitaker et al.\ are derived from medium-band photometry using \cite{bc03} models that assume a \cite{chabrier03} initial mass function (IMF), solar metallicity, and the \cite{calzetti00} extinction law. The stellar masses that we use were estimated by  
\cite{borch06} using the COMBO-17 photometry in conjunction
with a template library derived using the PEGASE stellar population models \citep{fioc97} with solar metallicity and a \cite{kroupa01} IMF. The different IMFs used by Whitaker et al.\ and Borch et al.\  would only change the stellar masses by $\sim10$\%. Moreover, the stellar masses derived in these works are quantitatively consistent with those derived using a simple colour€"-based stellar mass-to-light ratio \citep{bell03}. We are therefore confident that the stellar masses used here can be directly compared, and that any systematic differences will not affect our conclusions. The SFRs estimated by \cite{whitaker12} are based on UV+IR indicators, and are therefore expected to be on average $\sim0.3$\,dex higher, than the \Ha-based ones. This has to be taken into account when comparing our data with that of Whitaker et al. Nevertheless, even if we consider such offset and any likely systematic uncertainties, it is clear that most of our \Ha-detected galaxies would still fall clearly below the field SFR-mass relation, consistent with the differences found between our \Ha LF and the field one (Fig.~\ref{lumfunc}).  This confirms that many of the star-forming galaxies located in the centres of the A901/2 clusters appear to have their SFR significantly suppressed relative to the field.  This suppression seems to occur at all stellar masses. 

At first sight, this seems to contradict previous results which suggest that the average SFR in SF galaxies remains roughly constant with environment, and it is only the fraction of SF galaxies that changes \citep{balogh04,verdugo08,poggianti08,bamford08}. However, this contradiction is only apparent. The relative fraction of SF and non-SF galaxies depends on the depth of the survey. Because we are able to detect very low \Ha fluxes (and thus very low levels of star formation), our sample contains \Ha-emitting galaxies that would have been missed by most other surveys. Therefore, strongly suppressed galaxies still appear as \Ha emitters in our sample. Cluster galaxies with strong star formation (exemplified by the blue stars in Fig.~\ref{Mass_SFR}) have levels of SF comparable with those found in the field. If only galaxies with strong star formation are detected in a survey, it would appear that their SF has not been suppressed. Since galaxies with very low levels of star formation would not have been detected in \Ha in most other surveys, they would have been (erroneously) classified as non-SF and one would conclude that only the relative fraction of SF and non-SF galaxies has been affected by the environment.

\cite{wolf09} reached the same conclusion based on restframe UV and $24\mu$-data of Abell 901/2. They found a large population of lower-SFR galaxies, mostly red spiral galaxies seemingly in transition from normally star-forming galaxies to quiescent ones. This population had higher FIR luminosities than expected from their broad-band optical colour alone, suggesting that the specific SFR of the average red spiral at a given stellar mass was only three times lower than the SSFR of the average blue spiral at similar mass. An earlier study by \cite{wolf05} had already revealed this population from optical photometry alone, using mediumband imaging to discriminate between red colours due to high age in the absence of dust and those due to moderate age in the presence of moderate amounts of dust, thus differentiating between what they called 'old red' galaxies and 'dusty red' galaxies. The average specific SFR estimated from \OII emission in their 'dusty red' population was four times lower than that in blue galaxies. While this paper investigates only a small part of the same cluster, we find a similar transition population in SFRs estimated from \Ha now. We are thus suggesting that this cluster contains a rich population of moderately star-forming galaxies filling the previously considered gap between non-star-forming and star-forming galaxies. This view is now supported by evidence from a range of tracers, from UV and 24$\mu$, from \OII and H$\alpha$-emission, and from a detailed analysis of the optical SED.
A detailed study of the environmental dependence of the SF for the full galaxy sample will be presented in a subsequent paper.

\section[]{Summary and Conclusions}
\label{conclusions}

In this paper we have introduced the OMEGA (Osiris Mapping of Emission-line Galaxies in A901/2) survey.
The ultimate goal of this project is to study star formation and AGN activity across a broad range of environments at a single redshift.
Using the tuneable-filter mode of the OSIRIS instrument on the 10.4m GTC we have targeted \Ha and \NII emission lines over a $\sim \! 0.5 \times 0.5\;\mathrm{deg}^2$ region covering the $z \sim 0.167$ multi-cluster system containing Abell 901 and 902. 

Based on observations of 2 out of the 20 fields covered by this survey, this paper outlines the data reduction procedures followed and describes the analysis we have developed in order to reliably measure the fluxes and equivalent widths of the \Ha and \NII lines. We present the quantitative criteria used to define a robust sample of emission-line galaxies and analyse its properties in terms of detection efficiency, detection limits, and completeness. This paper also presents some examples of the kind of science that can be done with the survey.  The scientific results we discuss are based on only 10\% of the surveyed area, corresponding to the highest density regions of the A901/2 structure. An analysis of the full survey is left for future papers, but interesting results are already emerging. 

\begin{itemize}

\item In these two fields we detect 124 ELGs down to \Ha fluxes  $\sim1.3\times10^{-17}\,$erg$\,$cm$^{-2}$s$^{-1}$, corresponding to \Ha luminosities $\sim10^{39}\,$erg$\,$s$^{-1}$ at the distance of the A901/2 system. This allows us to probe integrated SFRs as low as $\sim \! 0.02 \; M_{\sun}\,\mathrm{yr}^{-1}$. We estimate that we are able to detect \Ha emission in $\sim40$\% of the cluster members with stellar masses above $\sim10^9M_{\sun}$, and $\sim60$\% of the galaxies expected to be forming stars according to their SEDs in the same mass range.

\item Using the WHAN diagram \citep{cid10,cid11}, which plots \WHa vs.\ the \NII/\Ha ratio, we are able to identify 10 bona-fide AGN in the Abell 901 and 902 clusters. The AGN hosts tend to be brighter and more massive than the star-forming galaxies. Moreover, the AGN host galaxies usually have earlier morphologies than the general star-forming galaxy population.

\item The emission-line galaxies we detect, both star-forming and AGN, tend to avoid the central highest-density regions of the clusters. 

\item The \Ha luminosity function of the galaxies in the densest regions of Abell 901/2 contains brighter \Ha emitters than that of more massive and richer clusters at comparable redshifts (e.g., Abell 1689 at $z=0.18$). However, our sample contains a lower proportion of very luminous \Ha emitters than the field galaxy population at similar redshifts. This indicates a progressive suppression of the star-formation activity in galaxies when we move from the field to clusters of increasing richness/mass.  

\item The SFR--stellar mass relation we observe in the densest regions of A901/2 falls well below the relation for field galaxies found by \cite{whitaker12} at similar redshifts. In other words, the average specific SFR (or SFR per unit stellar mass) is significantly lower in these clusters that in the field, confirming the star-formation suppressing effect of the clusters. 

\item The SFRs determined from the \Ha luminosities correlate very well with the SFRs estimated from the rest-frame UV photometry, but are typically a factor of $\sim2$ higher. The strong correlation indicates that the UV light and \Ha trace roughly comparable star-formation timescales but the effect of extinction is significantly stronger in the UV than in the red. \Ha-determined SFRs also correlate well with those determined combining the UV and IR emission, but the offset is now reversed: the \Ha SFR estimates would miss heavily-obscured star formation and are thus a factor $\sim2$ smaller. Some of the difference could also be due to a declining SFR, since \Ha and the IR probe different star-formation timescales.

\end{itemize}

We have found that the OSIRIS tuneable filter is a highly effective instrument for mapping the emission lines of galaxies at $z \sim 0.2$. In this paper we have presented several good examples of its scientific capabilities and potential. These results have been obtained using only a small fraction of the complete OMEGA dataset and therefore provide only a partial view of the star-formation and AGN properties of the galaxies in the Abell 901/902 structure, biased towards the highest density regions. A more complete picture, covering a much broader range of environments, will emerge once the full sample is analysed in future papers. Moreover, here we have only considered the integrated emission-line properties of the galaxies. OMEGA also provides spatially-resolved emission-line maps which will be used to study how the spatial distribution of the star formation is affected by internal and environmental factors. In future papers we will also explore the information on the chemical composition of the ionised gas encoded in the \Ha/\NII emission-line ratio.

\section*{Acknowledgments}
Based on observations made with the Gran Telescopio Canarias (acquired through an ESO Large Programme), installed in the Observatorio del Roque de los Muchachos of the Instituto de Astrof\'isica de Canarias, in the island of La Palma. Antonio Cabrera Lavers is thanked for invaluable support during the observations and Alessandro Ederoclite for advice with the use of OOPs. 
This work has made use of The University of Nottingham HPC facility, "Minerva".
ACS acknowledges funding from a CNPq, BJT-A fellowship (400857/2014-6). 
SPB and MEG gratefully acknowledge the receipt of an STFC Advanced Fellowship. AB is funded by the Austrian Science Foundation FWF (grant P23946-N16).

\label{lastpage}

\end{document}